\documentclass[a4paper,11pt]{article}
\pdfoutput=1 % if your are submitting a pdflatex (i.e. if you have
\usepackage{jheppub} % for details on the use of the package, please
\usepackage[T1]{fontenc} % if needed

\usepackage[utf8]{inputenc}
\usepackage{amssymb}
\usepackage{amsmath}
\usepackage{amsfonts}
\usepackage{graphicx}
\usepackage{color}
\usepackage{xspace}
\usepackage{comment}
\usepackage{hyperref}
\usepackage[normalem]{ulem}
\usepackage[section]{placeins}
\usepackage{afterpage}
\usepackage{slashed}
\usepackage{enumerate}
\usepackage{enumitem}
\usepackage{caption}
\usepackage{subfigure}
\usepackage{float}
\usepackage{ulem}
\usepackage{verbatim}
\usepackage{multirow,rotating}
\usepackage[dvipsnames]{xcolor}
\usepackage{indentfirst}
\makeatletter
\def\@fpheader{} 
\makeatother

\definecolor{dcolour}{rgb}{.5, .5, .5}
\def\gsim{\raise0.3ex\hbox{$\;>$\kern-0.75em\raise-1.1ex\hbox{$\sim\;$}}}
\def\lsim{\raise0.3ex\hbox{$\;<$\kern-0.75em\raise-1.1ex\hbox{$\sim\;$}}}
\def\gsim{\raise0.3ex\hbox{$\;>$\kern-0.75em\raise-1.1ex\hbox{$\sim\;$}}}
\def\lsim{\raise0.3ex\hbox{$\;<$\kern-0.75em\raise-1.1ex\hbox{$\sim\;$}}}

\newcommand{\ba}[1]{\begin{eqnarray} \label{(#1)}}
\newcommand{\ea}{\end{eqnarray}}

\newcommand{\iab}{\rm ab^{-1}}

\newcommand{\met}{\slashed{E}_T}

\newcommand{\GeV}{\,{\mathrm{GeV}}}
\newcommand{\TeV}{\,{\mathrm{TeV}}}

\title{\boldmath Discovery prospects for photophobic axion-like particles in the $WWjj$ final state at the High-Luminosity LHC}

\author[a, 1]{Jiaojiao Feng\note{Corresponding author.}}
\author[a, 1]{, Ying-nan Mao}
\author[a, 1]{and Kechen Wang}

\affiliation[a]{Department of Physics, School of Physics and Mechanics, Wuhan University of Technology,\\430070 Wuhan, Hubei, China}

\emailAdd{jiaojiao.feng@whut.edu.cn}
\emailAdd{ynmao@whut.edu.cn}
\emailAdd{kechen.wang@whut.edu.cn}

\abstract{
We evaluate discovery prospects for \emph{photophobic} axion-like particles (ALPs) in the $WWjj$ final state at the High-Luminosity LHC (HL-LHC; $\sqrt{s}=14~\mathrm{TeV}$, $\mathcal{L}=3~\mathrm{ab}^{-1}$). In the photophobic limit ($g_{a\gamma\gamma}=0$), ALPs couple to electroweak gauge bosons and are produced in association with two jets ($pp\to jj\,a$) via both $s$-channel electroweak exchange and vector-boson–fusion (VBF)–like topologies, followed by $a\to W^+W^-$ decay. We target the different-flavour dilepton mode $W^+W^-\!\to e^\pm\mu^\mp\nu \bar{\nu}$ with two jets and moderate missing transverse momentum. 
The analysis employs a two-step strategy: an initial preselection that defines the signal-like final state, followed by a multivariate analysis (MVA) trained on dijet and dileptonic-$WW$ kinematics from \emph{both} the $s$-channel and VBF-like production mechanisms to separate signal from background; the MVA threshold is optimized independently at each $m_a$.
We present $2\sigma$ and $5\sigma$ discovery sensitivities for the ALP–$W$ coupling $g_{aWW}$ across $170$–$4000~\mathrm{GeV}$ mass range. 
For $260~\mathrm{GeV}\le m_a\le1500~\mathrm{GeV}$, the $2\sigma$ ($5\sigma$) sensitivity is approximately flat around $0.61\,(0.76)~\mathrm{TeV}^{-1}$. We also report model-independent discovery thresholds for the fiducial quantity $\sigma(pp\to jj\,a)\times\mathrm{Br}(a\to W^+W^-)$ over the same mass range to enable reinterpretation for other models.
These results indicate that the $WWjj$ topology offers competitive and complementary sensitivity to heavy photophobic ALPs at the HL-LHC.
}

\begin{document}
\maketitle
\flushbottom

\section{Introduction}
\label{sec:intro}

Quantum chromodynamics (QCD) allows for a CP-violating $\theta$ term, yet no corresponding effect has been seen in measurements of the neutron electric dipole moment~\cite{Baker:2006ts,Abel:2020pzs}. This long–standing ``strong CP problem'' can be resolved by enlarging the Standard Model (SM) with a global $\mathrm{U}(1)_{\rm PQ}$ symmetry, known as the Peccei-Quinn symmetry, whose spontaneous breaking generates a pseudo-Nambu–Goldstone boson, the axion~\cite{Peccei:1977hh,Peccei:1977ur,Weinberg:1977ma,Wilczek:1977pj}. In traditional QCD-axion models, the axion mass and its couplings are tightly linked to the symmetry-breaking scale~\cite{Weinberg:1977ma,Wilczek:1977pj,Kim:1979if,Shifman:1979if,Zhitnitsky:1980tq,Dine:1981rt,Quevillon:2019zrd,Hook:2019qoh}. 

In many extensions of the SM, however, pseudo-scalar states with axion-like couplings appear, whose masses and interaction strengths are not fixed by the strong CP solution. These states are collectively referred to as axion-like particles (ALPs)~\cite{Svrcek:2006yi,Irastorza:2018dyq,DiLuzio:2020wdo,Choi:2020rgn}. The decoupling of mass and couplings opens a large parameter space and has led to extensive search programs, ranging from astrophysical and cosmological probes to intensity-frontier experiments and high-energy colliders~\cite{Jaeckel:2015jla,Galanti:2022ijh,Liu:2017zdh,Bauer:2017ris,Dolan:2017osp,Bauer:2018uxu,dEnterria:2021ljz,Agrawal:2021dbo,Zhang:2021sio,Tian:2022bmm,Ghebretinsaea:2022xnj,deGiorgi:2022oks,Wang:2022ock,Schafer:2022shi,Cheung:2023nzg,Antel:2023hkf,Dev:2023hax,Biswas:2023ksj,Yue:2023mew,Yue:2023mjm,Balkin:2023gya,Lu:2024fxs,Qiu:2024muo,Cheung:2024qve,Esser:2024pnc,Wang:2024pqa,Marcos:2024yfm,Yang:2024dic,Li:2024zhw,Inan:2025bdw,Yue:2025ksr,Ai:2025cpj,Wang:2025ncc,Bao:2025tqs,Alda:2025uwo,Wang:2025vhy,Bedi:2025hbz,Figliolia:2025dtw,Yue:2025wkn}.

At colliders, ALPs can interact with SM gauge bosons and/or fermions and can be probed either as on-shell resonances or through off-shell exchange in high-energy processes. 
A large body of experimental work has already constrained the ALP coupling to photons, $g_{a\gamma\gamma}$, using $e^+e^-$ data and heavy-ion or proton–proton collisions, in particular via diphoton or light-by-light signatures~\cite{CMS:2018erd,Belle-II:2020jti,ATLAS:2020hii,ATLAS:2022abz,BESIII:2022rzz,BESIII:2024hdv,CMS:2024bnt,ALPlimits}. 
The resulting bounds are strong over a wide mass range and motivate scenarios in which $g_{a\gamma\gamma}$ is tuned to be small while couplings to the electroweak gauge bosons remain sizeable. 
In such \emph{photophobic} constructions, the $a\gamma\gamma$ vertex is suppressed but interactions with $W^\pm$, $Z$ and $\gamma Z$ can persist~\cite{Craig:2018kne,Fonseca:2018xzp,Hook:2016mqo}, leading to collider signatures that differ qualitatively from generic diphoton ALP searches.

Heavy photophobic ALPs at the LHC have been explored in several ways.  
One line of work has reinterpreted SM triboson measurements at $\sqrt{s}=8~\TeV$ as signals of ALP-mediated processes, deriving limits up to a few hundred GeV in mass~\cite{Craig:2018kne}. 
Other studies have mapped nonresonant ALP exchange onto vector-boson-scattering topologies~\cite{Bonilla:2022pxu}, or recast Run-2 analyses targeting $W^\pm W^\pm W^\mp$ and $Z\gamma$ final states to constrain resonant ALP production and decay into electroweak boson pairs~\cite{Aiko:2024xiv}. 
An overview of these and related collider bounds can be found in Ref.~\cite{Ding:2024djo}. 
Since the original analyses were optimized for SM processes rather than for on-shell ALP production, differences in acceptance, kinematic selections and background control often make the recast limits conservative and non-uniform across the parameter space.

More recently, dedicated projections have been carried out specifically for photophobic ALPs at the High-Luminosity LHC (HL-LHC)~\cite{CERN-2019-007}.
In Ref.~\cite{Ding:2024djo}, a detector-level study of $pp\to jj\,a(\to Z\gamma)$ was performed, including both $s$-channel and vector-boson–fusion (VBF)–like production mechanisms, reconstructing the ALP resonance and using a multivariate analysis (MVA) to obtain discovery reaches on $g_{aWW}$ and on $\sigma(pp\to jj\,a)\times\mathrm{Br}(a\to Z\gamma)$ over $m_a\simeq(100$–$4000)~\GeV$ at $\sqrt{s}=14~\TeV$. 
In parallel, Ref.~\cite{Mao:2024kgx} proposed a tri-$W$ signature, $pp\to W^\pm X(\to W^+W^-)$, using a same-sign dimuon plus hadronic-$W$ final state and an MVA tailored for background rejection; with heavy photophobic ALPs used as a benchmark, that study obtained sensitivities well beyond earlier triboson recasts, while remaining applicable to more general neutral resonances coupled to $W$ bosons. 
These works highlight two important lessons: clean electroweak final states with sharp kinematic handles and multivariate discrimination can dramatically sharpen sensitivity, and photophobic ALP searches should exploit multiple channels with different couplings and systematics.

The present paper extends this program to the $WWjj$ topology. 
We consider photophobic ALPs that couple to the electroweak gauge bosons and are produced in association with two jets via both $s$-channel electroweak exchange and VBF-like topologies in $pp$ collisions, $pp\to jj\,a$, followed by $a\to W^+W^-$ decay. 
Both production mechanisms are simulated and analyzed on equal footing within a single selection-and-MVA framework. 
This strategy makes use of the full photophobic electroweak structure ($aWW$, $aZZ$, $a\gamma Z$) and allows the analysis to adapt to whichever kinematic pattern dominates: $s$-channel–like configurations are more important at lower $m_a$, while VBF-like events with forward jets and large rapidity gaps become increasingly relevant at high $m_a$. 
As a clean and experimentally robust final state, we focus on the different-flavour dilepton mode $W^+W^-\to e^\pm\mu^\mp\nu\bar{\nu}$ accompanied by at least two jets, which efficiently suppresses Drell–Yan and multijet backgrounds. 
Our study provides projected discovery sensitivities on both the coupling $g_{aWW}$ and the fiducial quantity $\sigma(pp\to jj\,a)\times\mathrm{Br}(a\to W^+W^-)$ 
at the HL-LHC with $\sqrt{s}=14~\TeV$ and $\mathcal{L}=3~\iab$.

Throughout this work, unless stated otherwise, we impose the photophobic condition $g_{a\gamma\gamma}=0$, include both $s$-channel and VBF-like production of $pp\to jj\,a$ with $a\to W^+W^-$ decay, and optimize the boosted decision tree (BDT) threshold \emph{independently} for each assumed ALP mass. The quoted significances are purely statistical (systematic uncertainties are not included; cf. Sec.~\ref{sec:results}). 

We note that our HL-LHC projection in the $pp\!\to\!jj\,a(\to W^+W^-)$ topology should be read in the context of several recent and complementary studies, including the $jj\,a(\to Z\gamma)$ channel~\cite{Ding:2024djo}, the tri-$W$ strategy $pp\!\to\!W^\pm X(\to W^+W^-)$~\cite{Mao:2024kgx}, and related Run-1/2 recasts~\cite{Craig:2018kne,Aiko:2024xiv,Bonilla:2022pxu}. 
Relative to the tri-$W$ strategy~\cite{Mao:2024kgx}, our $WWjj$ analysis typically achieves a stronger reach across the full mass range, with a particularly pronounced advantage for $m_a\gtrsim 700~\mathrm{GeV}$, owing to the VBF-like production and kinematics. 
By contrast, $jj\,a(\to Z\gamma)$~\cite{Ding:2024djo} attains a stronger reach overall—benefiting from a narrow $Z\gamma$ resonance and cleaner backgrounds—yet $jj\,a(\to W^+W^-)$ remains complementary: it \emph{directly} probes the $aWW$ vertex, thus enabling nontrivial cross-checks of the photophobic relations among $aWW$, $aZ\gamma$, and $aZZ$ couplings when combined with previous results~\cite{Ding:2024djo,Mao:2024kgx}; it is subject to largely orthogonal systematics and backgrounds relative to $Z\gamma$; and our model-independent thresholds in $\sigma(pp\!\to\!jj\,a)\times{\rm Br}(a\!\to\!W^+W^-)$ facilitate straightforward reinterpretation for other neutral resonances that couple to $W$ bosons~\cite{Mao:2024kgx}. 

This paper is organized as follows. 
Sec.~\ref{sec:theory} introduces the effective description of photophobic ALPs and their electroweak couplings. 
In Secs.~\ref{sec:signal} and~\ref{sec:SM_BKG}, we define the signal process and summarize the relevant SM backgrounds. 
The event selection, observable construction, and multivariate strategy are described in Sec.~\ref{sec:analysis}. 
Our main results are presented in Sec.~\ref{sec:results}, and we conclude in Sec.~\ref{sec:conc}. 
Additional distributions, BDT response shapes, and efficiency tables are collected in the appendices.

\section{Theory setup}
\label{sec:theory}

We work with an ALP whose interactions with the %\modified{
SM gauge bosons
%} 
arise solely through the $\mathrm{SU}(2)_{\rm L}$ and $\mathrm{U}(1)_{\rm Y}$ field strengths. At the level before electroweak symmetry breaking (EWSB), the relevant effective Lagrangian can be written as~\cite{Georgi:1986df}
\begin{equation}
\mathcal{L}_{\rm ALP} \supset
\frac{1}{2}\,\partial_{\mu}a\,\partial^{\mu}a
-\frac{1}{2}\,m_{a}^{2}\,a^{2}
-\frac{c_{\widetilde W}}{f_{a}}\,a\,W_{\mu\nu}^{b}\,\widetilde W^{b,\,\mu\nu}
-\frac{c_{\widetilde B}}{f_{a}}\,a\,B_{\mu\nu}\,\widetilde B^{\mu\nu},
\label{eq:L}
\end{equation}
where $a$ and $m_a$ denote the ALP field and its mass, and $f_a$ is the associated decay constant. The tensors $W_{\mu\nu}^{b}$ (with $b=1,2,3$) and $B_{\mu\nu}$ are the usual $\mathrm{SU}(2)_{\rm L}$ and $\mathrm{U}(1)_{\rm Y}$ field strengths, and the corresponding dual tensor is defined as $\widetilde X^{\mu\nu}\equiv \tfrac{1}{2}\epsilon^{\mu\nu\alpha\beta}X_{\alpha\beta}$ with $X=W^b,B$. The dimensionless coefficients $c_{\widetilde W}$ and $c_{\widetilde B}$ encode the strengths of the ALP couplings to the electroweak gauge sector.

After EWSB, it is convenient to express the interactions in the mass eigenstate basis for the gauge bosons. The terms involving neutral and charged electroweak bosons can be recast as
\begin{equation}
\mathcal{L}_{\rm int}
\supset -\frac{1}{4}\, g_{a\gamma\gamma}\, a\, F_{\mu\nu}\,\widetilde F^{\mu\nu}
  -\frac{1}{2}\, g_{aZ\gamma}\, a\, Z_{\mu\nu}\,\widetilde F^{\mu\nu}
  -\frac{1}{4}\, g_{aZZ}\, a\, Z_{\mu\nu}\,\widetilde Z^{\mu\nu}
  -\frac{1}{2}\, g_{aWW}\, a\, W_{\mu\nu}^{+}\,\widetilde W^{-\mu\nu},
\label{eq:L1}
\end{equation}
with the tree-level relations
\begin{align}
g_{a\gamma\gamma} &= \frac{4}{f_a}\,\big(s_\theta^{2}\,c_{\widetilde W}+c_\theta^{2}\,c_{\widetilde B}\big), \\
g_{aZZ} &= \frac{4}{f_a}\,\big(c_\theta^{2}\,c_{\widetilde W}+s_\theta^{2}\,c_{\widetilde B}\big), \\
g_{aZ\gamma} &= \frac{2}{f_a}\, s_{2\theta}\,\big(c_{\widetilde W}-c_{\widetilde B}\big), \\
g_{aWW} &= \frac{4}{f_a}\, c_{\widetilde W}.
\end{align}
Here $\theta$ is the weak mixing angle, and we define $s_\theta\equiv\sin\theta$, $c_\theta\equiv\cos\theta$, and $s_{2\theta}\equiv\sin 2\theta$ for simplify. Guided by the strong experimental constraints on the diphoton channel, we restrict ourselves to the photophobic case in which the $a\gamma\gamma$ coupling vanishes, $g_{a\gamma\gamma}=0$ (see, e.g.,~\cite{Craig:2018kne,Fonseca:2018xzp,Hook:2016mqo,Bonilla:2022pxu,Aiko:2024xiv,Ding:2024djo,Mao:2024kgx}). Imposing
\begin{equation}
s_\theta^{2} c_{\widetilde W}+c_\theta^{2} c_{\widetilde B}=0,
\quad\mathrm{or~equivalently}\quad
c_{\widetilde B} = - t_\theta^{2}\, c_{\widetilde W},
\end{equation}
with $t_\theta\equiv\tan\theta$, $g_{a\gamma\gamma}$ is removed at tree level and thus the remaining electroweak couplings correlate with each other. In particular, one finds 
\begin{align}
g_{aZZ} &= \big(1 - t_\theta^{2}\big)\, g_{aWW}, \\
g_{aZ\gamma} &= t_\theta\, g_{aWW}.
\end{align}
In the following, we use $m_a$ and $g_{aWW}$ as the primary parameters, with $g_{aZZ}$ and $g_{aZ\gamma}$ fixed by these relations throughout our phenomenological analysis of the photophobic scenario.

\section{Signal production}
\label{sec:signal}

In the photophobic limit $g_{a\gamma\gamma}=0$, the ALPs retains anomalous couplings to electroweak gauge bosons, notably $aWW$, $aZZ$, and $a\gamma Z$. We denote $V\in\{W,Z,\gamma\}$ and $V'\in\{W,Z\}$. Consequently, the ALP can be produced at proton colliders in association with one vector boson ($pp\to V' \,a$) through $s$-channel vector-boson exchange or with with two jets ($pp\to jj \,a$) via vector-boson-fusion (VBF)–like topologies, and it can decay as $a\to VV'$ with branching ratios varying with the ALP mass~\cite{Aiko:2023trb,Ding:2024djo}. For sufficiently large $m_a$, the $a\to W^+W^-$ mode dominates.

\begin{figure}[h]
\centering
\includegraphics[width=\linewidth]{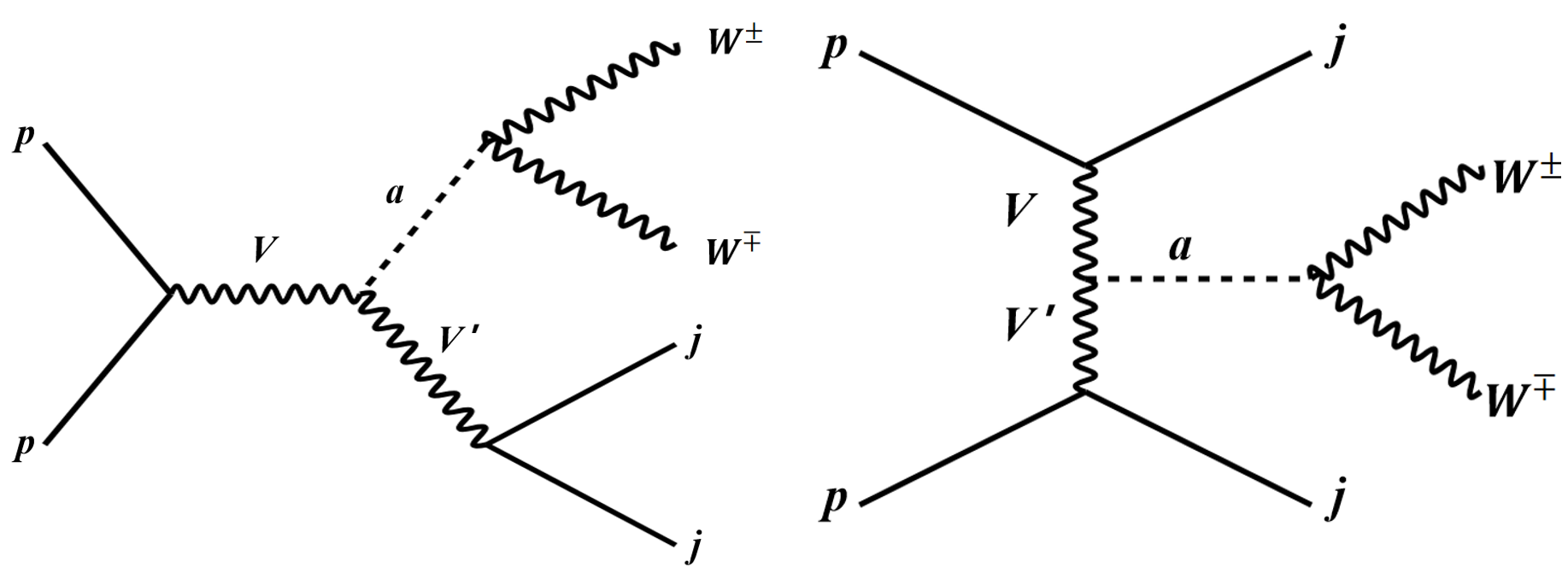}
\caption{Representative production mechanisms for heavy photophobic ALPs at $pp$ colliders: $s$-channel exchange (left) and VBF-like topology (right), yielding $pp\to jj\,a$ with $a$–$V$–$V'$ vertices ($aWW$, $aZZ$, $a\gamma Z$). The ALP subsequently decays to two $W$ bosons, defining the signal $pp\to jj\,a(\to W^\pm W^\mp)$ with $W^\pm\to \ell^\pm\bar{\nu}/\nu$.}
\label{fig:signal}
\end{figure}

As illustrated in Fig.~\ref{fig:signal}, we consider two representative production processes: (i) an $s$-channel electroweak-boson exchange yielding $pp\to jj\,a$, with the two jets predominantly arising from the decay of the associated vector boson $V'$; and (ii) an electroweak VBF-like topology in which the ALP is radiated off the $t$-channel gauge bosons. In both cases, the effective $a$–$V$–$V'$ vertices include $aWW$, $aZZ$, and $a\gamma Z$. We focus on the different-flavour dilepton final state, $W^+W^-\to e^\pm\mu^\mp\nu \bar{\nu}$, which suppresses multijet backgrounds and Drell–Yan $Z/\gamma^\ast\to \ell^+\ell^-$ contamination while retaining high signal efficiency. 

The VBF-like component is characterized by two forward tagging jets with large invariant mass $m_{jj}$ and sizeable pseudo-rapidity difference $|\Delta\eta_{jj}|$, together with reduced central hadronic activity;
whereas the $s$-channel contribution typically yields a more central dijet event with smaller rapidity separation and an invariant mass near the parent-boson mass when the jets arise from $V'\!\to jj$. 
Accordingly, the MVA in Sec.~\ref{sec:analysis} is trained on the inclusive (combined) signal and uses variables that are simultaneously sensitive to both topologies and kinematic observables that reconstruct the common decay $a\to W^+W^-\to e^\pm\mu^\mp\nu \bar{\nu}$ in the $e^\pm\mu^\mp+\slashed{E}_T$ final state.

Signal and background samples are generated with MadGraph5\_aMC@NLO~\cite{Alwall:2014hca} using the NNPDF2.3 parton distribution functions (PDF)~\cite{Ball:2012cx}. The ALP interactions are implemented via a \textsc{UFO} model based on the linear Lagrangian~\cite{Brivio:2017ije,Degrande:2011ua}. Parton showering and hadronization are performed with \textsc{PYTHIA}~8.3~\cite{Bierlich:2022pfr}, and the detector response is simulated with \textsc{Delphes}~3~\cite{deFavereau:2013fsa} using HL-LHC CMS detector cards. The HL-LHC center-of-mass energy is set to $\sqrt{s} = 14$ TeV (7 TeV per beam).
The hard-scattering process $pp\to jj\,a$ is generated with MadGraph5\_aMC@NLO, while the subsequent decays $a\to W^+W^-$ and $W^\pm\to \ell^\pm \nu / \bar{\nu}$ ($\ell=e,\mu$), yielding $a\to W^+(\to \ell^+\nu)\,W^-(\to \ell^-\bar\nu)\,jj$, are simulated with \textsc{PYTHIA}~8.3.

We generate signal samples at representative ALP masses
\[
m_a \in \{170,\ 185, \ 200,\ 230, \ 260,\ 350,\ 500,\ 750,\ 1000,\ 1500,\ 2000,\ 2500,\ 3000,\ 4000\}\ \mathrm{GeV}.
\]
At the HL-LHC, we generate $4.8$, $0.8$, and $4.0$ million events for the benchmark points $m_a=200,\,260,$ and $350~\mathrm{GeV}$, respectively; $0.4$ million events for each point with $m_a=170,\,185,\,230~\mathrm{GeV}$; $2.0$ million events for 
each point with
$m_a=500,\,750,\,4000~\mathrm{GeV}$; and $1.0$ million events for 
each point with
$m_a=1000,\,1500,\,2000,\,2500$, $3000~\mathrm{GeV}$. 
To remain inclusive at generator level and to match experimental conditions, we employ the following loose parton-level cuts in the MadGraph5 configuration card:
\begin{enumerate}[label*=(\roman*)]
\item Minimal transverse momentum of jets, photons and charged leptons:\\ $p_{\rm T}(j/\gamma/\ell^\pm) > 0.5~\mathrm{GeV}$;
\item Maximal pseudorapidity:\\ $|\eta(j)|<10$ for jets and $|\eta(\gamma/\ell^\pm)|<5$ for photons and charged leptons;
\item Minimal angular separation $\Delta R\equiv \sqrt{(\Delta\eta)^2+(\Delta\phi)^2}$ between any two objects:\\ $\Delta R>0.1$.
\end{enumerate}
These generator-level requirements are deliberately much looser than the analysis selection and are never applied as analysis cuts. 
They avoid generator-induced inefficiencies (i.e.\ an artificial suppression of soft/forward configurations) and only regulate infrared-sensitive regions for technical stability~\cite{Alwall:2014hca}. 
Event realism is achieved downstream: parton showering and hadronization model QCD radiation, and the detector simulation emulates experimental acceptance and resolution~\cite{Bierlich:2022pfr,deFavereau:2013fsa}. 
Accordingly, the fiducial phase space and physics acceptance are defined by the detector simulation together with the selection in Sec.~\ref{sec:analysis}, in line with LHC fiducial-measurement practice.

\begin{figure}[h]
\centering
\includegraphics[width=\linewidth]{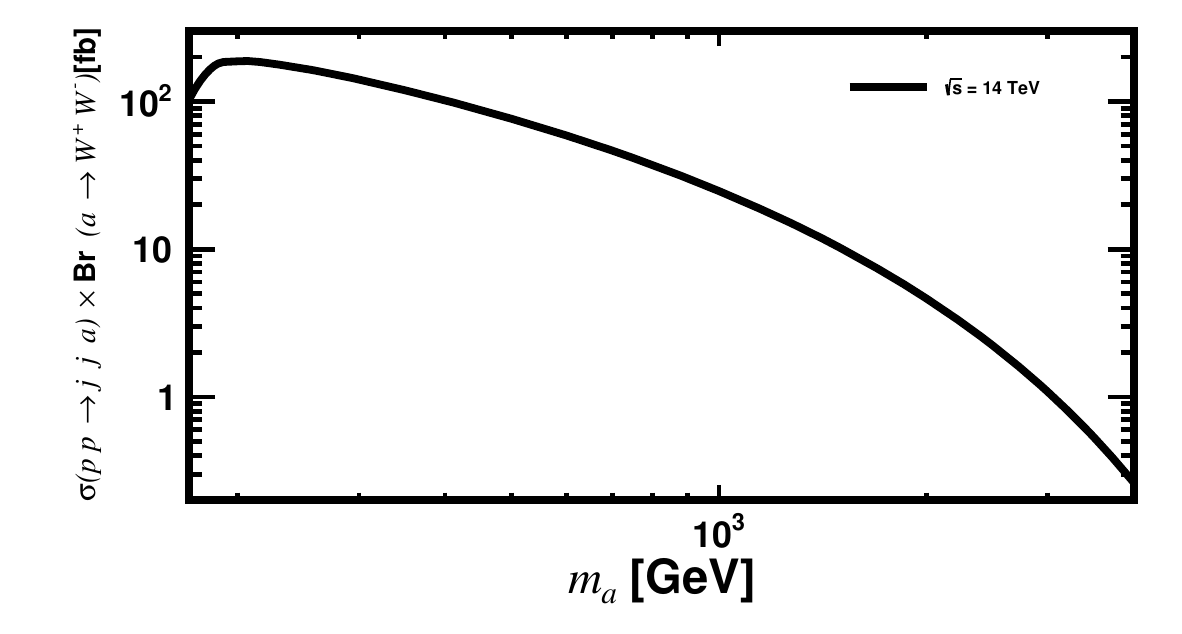}
\caption{Production cross section of $pp\to jj\,a(\to W^\pm W^\mp)$ at the HL-LHC ($\sqrt{s}=14~\mathrm{TeV}$) for $g_{a\gamma\gamma}=0$ and $g_{aWW}=1~\mathrm{TeV}^{-1}$ as a function of the ALP mass $m_a$.}
\label{fig:crosssection}
\end{figure}

Fig.~\ref{fig:crosssection} shows the production cross section for $pp\to jj\,a(\to W^\pm W^\mp)$ at $\sqrt{s}=14~\mathrm{TeV}$ under the benchmark couplings $g_{a\gamma\gamma}=0$ and $g_{aWW}=1~\mathrm{TeV}^{-1}$. The two-body $WW$ decay channel opens rapidly once $m_a\gtrsim 2m_W\simeq160~\mathrm{GeV}$, causing a steep rise in $\mathrm{Br}(a\to W^+W^-)$; for example, at $m_a=185~\mathrm{GeV}$, the branching fraction reaches $\sim49\%$~\cite{Aiko:2023trb,Ding:2024djo}. Meanwhile, the inclusive production cross section $\sigma(pp\to jj\,a)$ decreases with increasing $m_a$ due to diminishing PDFs, such that $\sigma(pp\to jj\,a)\times\mathrm{Br}(a\to W^+W^-)$ peaks around $m_a\sim200~\mathrm{GeV}$.

\section{Background processes}
\label{sec:SM_BKG}

The signal targets two \emph{oppositely charged, different-flavour} leptons, at least two jets, and moderate missing transverse momentum,
i.e.,\ $e^\pm\mu^\mp+jj+\slashed{E}_T$.
In the $e\mu$ channel, Drell--Yan contamination is strongly suppressed; the dominant backgrounds are well known to include diboson and top-quark production, as well as processes with nonprompt or misidentified leptons. We group them by physics origin and reducibility at detector level.

\paragraph{Irreducible (same visible final state)}
\begin{itemize}
  \item \textbf{$W^+W^-jj$ (QCD+EW).} Inclusive $WW$ production accompanied by jets, dominated by $q\bar q\!\to WW$ with additional QCD radiation~\cite{CMS:WW136emu2024}. 
  %and a smaller loop-induced $gg\!\to WW$ component. 
  After leptonic $W$ decays, the visible final state coincides with the signal ($e^\pm\mu^\mp+jj+\slashed{E}_T$).
  \item \textbf{$t\bar t$ (dileptonic).} When both $W$ bosons decay leptonically, $t\bar t\!\to\!\ell^+\nu\,\ell^-\bar\nu\,b\bar b$ can be indistinguishable from the signal up to global kinematics if $b$ jets are 
  not 
  correctly tagged
  ; it is thus effectively irreducible in this channel.
\end{itemize}

\paragraph{Reducible (lost/misidentified objects)}
\begin{itemize}
  \item \textbf{$WZjj$, $ZZjj$.} 
  Either 
  when one prompt lepton is not reconstructed or fails identification/isolation, or when $Z\!\to\!\tau\tau$ produces a pair $e\mu$.
  \item \textbf{$W^\pm jj$.} Contributes via one prompt lepton from the $W$ and one nonprompt or misidentified lepton (heavy-flavour decays, hadrons mis-identification, photon conversions).
  \item \textbf{$Zjj$.} Suppressed in $e\mu$ but nonzero through a misidentified lepton, a lost lepton plus a nonprompt lepton, or via $Z\!\to\!\tau\tau\!\to\!e\mu+X$. 
\end{itemize}

The $pp \to W^+ W^- jj$ background is generated at matrix-element level with MadGraph5\_aMC@NLO, while the subsequent decays $W \to \ell^\pm \nu_\ell$ ($\ell = e,\mu$) and parton showering are modeled with PYTHIA~8.3. 
The $t\bar{t}$, $WZjj$, and $ZZjj$ samples are likewise produced with MadGraph5\_aMC@NLO and interfaced to PYTHIA~8.3, which handles all kinematically allowed $W$- and $Z$-boson decays. 
Single-boson processes $W^\pm(\to \ell^\pm \nu_\ell)jj$ and $Z(\to \ell^+\ell^-)jj$ with $\ell = e,\mu$ are generated fully differentially in MadGraph5\_aMC@NLO, keeping the leptonic $W$ and $Z$ decays explicit in the matrix element.

\begin{table}[h]
\centering
\caption{Generator-level inclusive production cross sections of background processes at the HL-LHC ($\sqrt{s}=14~\mathrm{TeV}$). Values are prior to analysis selections; the dominant $Wjj$ and $Zjj$ rates will be strongly reduced by the dilepton requirements discussed in Sec.~\ref{sec:analysis}.}
\label{tab:bkg_xsec}
\begin{tabular}{lc}
\hline\hline
Background & $\sigma~[\mathrm{pb}]$ \\
\hline
$W^+ (\to \ell^+ \nu) \, W^-(\to \ell^- \bar{\nu}) jj$
& $3.62\times 10^{1}$ \\
$WZjj$ & $3.44\times 10^{2}$ \\
$ZZjj$ & $1.03\times 10^{2}$ \\
$W^\pm(\to \ell^\pm \nu_\ell/ \bar{\nu}_\ell)jj$ & $2.42\times 10^{5}$ \\
$Z(\to \ell^+\ell^-)jj$ & $2.21\times 10^{4}$ \\
$t\bar t$ & $5.97\times 10^{2}$ \\
\hline\hline
\end{tabular}
\end{table}

The expected number of events scales with luminosity as
\begin{equation}
N = \sigma_{\rm pro}\times\mathcal{L}\times \varepsilon_{\rm sel},
\label{eq:event_yield}
\end{equation}
where $\sigma_{\rm pro}$ is the production cross section, $\mathcal{L}$ is the integrated luminosity, and $\varepsilon_{\rm sel}$ is the 
net acceptance efficiency multiplied by the efficiencies from object reconstruction and analysis selections. 
Table~\ref{tab:bkg_xsec} shows generator-level inclusive production cross sections of background processes at the HL-LHC ($\sqrt{s}=14~\mathrm{TeV}$).
In practice, although the $Wjj$ and $Zjj$ processes dominate at generator level (see Table~\ref{tab:bkg_xsec}), their contribution after the dilepton+$\met$ selections is strongly reduced, while $WWjj$ and $t\bar t$ processes remain the leading backgrounds in the $e\mu$ channel~\cite{CMS:WW136emu2024}.

\section{Data analysis} 
\label{sec:analysis}

\subsection{Preselection}
\label{subsec:presel}

We target final states with two oppositely charged, different-flavour leptons, at least two jets, and moderate missing transverse momentum. Throughout, objects are ordered by transverse momentum, and the leading (subleading) jet is denoted by $j_1$ ($j_2$). The following preselection is applied to select the $e^\pm\mu^\mp+jj+\met$ final state and suppress backgrounds at the first stage:
\begin{enumerate}[label*=(\roman*)]
\item Exactly two leptons: $N(\ell)=2$;
\item Lepton transverse momenta: $p_T(\ell)\ge 10~\mathrm{GeV}$ for both leptons;
\item Different flavour: exactly one electron and one muon, $N(e)=1$ and $N(\mu)=1$;
\item Opposite charge: $Q(e)+Q(\mu)=0$;
\item Jet multiplicity: at least two jets, $N(j)\ge 2$;
\item Leading-jet thresholds: $p_T(j_1)\ge 30~\mathrm{GeV}$ and $p_T(j_2)\ge 30~\mathrm{GeV}$;
\item Top-quark suppression: zero $b$-tagged jets, $N(j_b)=0$.
\end{enumerate}
This opposite-sign $e\mu$ requirement strongly reduces Drell--Yan backgrounds, while the $b$-jet veto suppresses top-quark processes, in line with LHC $WW\to e^\pm\nu\mu^\mp \bar{\nu}$ measurements~\cite{CMS:WW136emu2024}. 

\begin{table}[h]
\centering
\caption{Preselection efficiencies for signal at representative $m_a$ values and for background processes at the HL-LHC with $\sqrt{s}=14~\mathrm{TeV}$. Numbers are efficiencies after applying criteria (i)--(vii).}
\label{tab:preselection14TeV}
\scalebox{0.8}{
\begin{tabular}{cc}
\hline\hline
\begin{tabular}{c}
$m_a~[\mathrm{GeV}]$ \\
Efficiency 
\end{tabular} &
\begin{tabular}{ccccccc}
170 & 185 & 200 & 230 & 260 & 350 & 500 \\
~~~$0.150$~~~ & ~~~$0.151$~~~ & ~~~$0.152$~~~ & ~~~$0.156$~~~ & ~~~$0.158$~~~ & ~~~$0.165$~~~ & ~~~$0.175$~~~
\end{tabular}\\
\hline
\begin{tabular}{c}
$m_a~[\mathrm{GeV}]$ \\
Efficiency
\end{tabular} & 
\begin{tabular}{ccccccc}
750 & 1000 & 1500 & 2000 & 2500 & 3000 & 4000 \\
~~~$0.186$~~~ & ~~~$0.194$~~~ & ~~~$0.201$~~~ & ~~~$0.204$~~~ & ~~~$0.205$~~~ & ~~~$0.201$~~~ & ~~~$0.191$~~~ 
\end{tabular}\\
\hline
\begin{tabular}{c}
Background \\
Efficiency
\end{tabular} &
\begin{tabular}{cccccc}
$W^+ (\to \ell^+ \nu) \, W^-(\to \ell^- \bar{\nu}) jj$ & $WZjj$ & $ZZjj$ & $W^\pm(\to \ell^\pm \nu_\ell/ \bar{\nu}_\ell)jj$ & $Z(\to \ell^+\ell^-)jj$ & $t\bar t$ \\
$2.65\times 10^{-2}$ & $4.64\times 10^{-4}$ & $2.82\times 10^{-4}$ & $1.33\times 10^{-6}$ & $3.16\times 10^{-6}$ & $1.21\times 10^{-3}$ 
\end{tabular}\\
\hline\hline
\end{tabular}
}
\end{table}

As summarized in Table~\ref{tab:preselection14TeV}, the $Wjj$ and $Zjj$ processes have the lowest preselection efficiencies because of the tight dilepton requirement. 
Diboson $WZjj$ and $ZZjj$ remain as subleading reducible backgrounds primarily through lost-lepton or $\tau$-lepton decays, while $WWjj$ and $t\bar t$ are effectively irreducible up to global kinematics; the latter is also strongly reduced by the $b$-tag veto.

\begin{table}[h]
\centering
\caption{Event yields after sequentially applying preselection criteria (i)--(vii) for the signal with benchmark $m_a=750~\mathrm{GeV}$ and for background processes at the HL-LHC with $\sqrt{s}=14~\mathrm{TeV}$ and $\mathcal{L}=3~\mathrm{ab}^{-1}$.}
\label{tab:presel}
\scalebox{0.7}{
\begin{tabular}{lccccccccc}
\hline\hline
HL-LHC & initial & (i) & (ii) & (iii) & (iv) & (v) & (vi) & (vii) \\
\hline
Signal & $1.00\times 10^{6}$ & $5.53\times 10^{5}$ & $5.36\times 10^{5}$ & $2.65\times 10^{5}$ & $2.65\times 10^{5}$ & $2.64\times 10^{5}$ & $2.25\times 10^{5}$ & $1.87\times 10^{5}$ \\
$W^+ (\to \ell^+ \nu) \, W^-(\to \ell^- \bar{\nu}) jj$
& $1.09\times 10^{8}$ & $3.49\times 10^{7}$ & $3.36\times 10^{7}$ & $1.65\times 10^{7}$ & $1.65\times 10^{7}$ & $1.51\times 10^{7}$ & $3.41\times 10^{6}$ & $2.88\times 10^{6}$ \\
$WZjj$ & $1.03\times 10^{9}$ & $2.71\times 10^{7}$ & $2.53\times 10^{7}$ & $2.51\times 10^{6}$ & $1.36\times 10^{6}$ & $1.31\times 10^{6}$ & $5.77\times 10^{5}$ & $4.78\times 10^{5}$ \\
$ZZjj$ & $3.10\times 10^{8}$ & $1.40\times 10^{7}$ & $1.37\times 10^{7}$ & $3.14\times 10^{5}$ & $2.23\times 10^{5}$ & $2.18\times 10^{5}$ & $1.15\times 10^{5}$ & $8.75\times 10^{4}$ \\
$W^\pm(\to \ell^\pm \nu_\ell/\bar\nu_\ell)jj$ & $7.26\times 10^{11}$ & $7.78\times 10^{8}$ & $3.10\times 10^{7}$ & $1.37\times 10^{7}$ & $7.50\times 10^{6}$ & $6.87\times 10^{6}$ & $1.28\times 10^{6}$ & $9.64\times 10^{5}$ \\
$Z(\to \ell^+\ell^-)jj$ & $6.62\times 10^{10}$ & $1.81\times 10^{10}$ & $1.78\times 10^{10}$ & $2.10\times 10^{6}$ & $1.06\times 10^{6}$ & $9.88\times 10^{5}$ & $2.71\times 10^{5}$ & $2.09\times 10^{5}$ \\
$t\bar t$ & $1.79\times 10^{9}$ & $4.35\times 10^{7}$ & $3.78\times 10^{7}$ & $1.85\times 10^{7}$ & $1.84\times 10^{7}$ & $1.84\times 10^{7}$ & $1.53\times 10^{7}$ & $2.16\times 10^{6}$ \\
\hline\hline
\end{tabular}
}
\end{table}

Table~\ref{tab:presel} shows the event yields after sequentially applying preselection criteria (i)--(vii) for the signal with benchmark $m_a=750~\mathrm{GeV}$ and for background processes at the HL-LHC with $\sqrt{s}=14~\mathrm{TeV}$ and $\mathcal{L}=3~\mathrm{ab}^{-1}$.
With the opposite-sign $e\mu$ requirement, the large inclusive $Wjj$ and $Zjj$ rates are greatly reduced, while $WWjj$ and $t\bar t$ remain the leading backgrounds in the $e\mu$ channel~\cite{CMS:WW136emu2024}.

\subsection{Multivariate analysis}
\label{subsec:multi}

After the preselection, we employ a multivariate analysis (MVA) to further separate signal from background. The input observables are:

\begin{enumerate}[label*=(\roman*)]
\item Object kinematics: energy, transverse momentum, pseudorapidity and azimuth of the final-state objects,
$E(O)$, $p_T(O)$, $\eta(O)$, $\phi(O)$ with $O=\mu,\,e,\,j_1,\,j_2$.
\item Missing transverse momentum (magnitude and azimuth): $\met$ and $\phi(\met)$.
\item Among all jet pairs, the pair with the minimal angular separation: $\Delta R (j, j')_{\rm min}$ and the corresponding dijet mass $m(j+j')_{ {\rm min} \Delta R}$, where $\Delta R = \sqrt{(\Delta \eta)^2 + (\Delta \varphi)^2}$.
\item Among all jet pairs, the pair with the maximal pseudorapidity separation: $\Delta \eta(j, j')_{\rm max}$ and the corresponding dijet mass $m(j + j')_{ {\rm max} \Delta \eta}$.
\item the invariant mass of first two leading jets and the jet multiplicity: $m(j_{1}+j_{2})$ and $N(j)$.
\item Angular separations between key objects:
$\Delta R(j_1,j_2)$, $\Delta\eta(j_1,j_2)$, $\Delta R(e,\mu)$, $\Delta R(\mu,j_1)$, $\Delta R(\mu,j_2)$, $\Delta R(e,j_1)$, $\Delta R(e,j_2)$.
\item Transverse masses of leptons and $\met$: $m_{\rm T}(\mu + \met)$, $m_{\rm T}(e + \met)$, and  $m_{\rm T}(\mu + e + \met)$. 
Here, the transverse mass $m_{\mathrm T}$ of a visible system and the missing transverse momentum is defined as
$
m_{\mathrm T}^2
= \big(E_{\mathrm T}^{\rm vis} + \met \big)^2
 - \bigl|\vec{p}_{\mathrm T}^{\,\rm vis} + \vec{p}_{\mathrm T}^{\,\rm miss}\bigr|^2 ,
$
where $E_{\mathrm T}^{\rm vis} = \sqrt{|\vec{p}_{\mathrm T}^{\,\rm vis}|^2 + (m^{\rm vis})^2}$ is the transverse energy of the visible system and 
$\met = |\vec{p}_{\mathrm T}^{\,\rm miss}|$ assumes a massless invisible system.
\end{enumerate}

Observables (iv)–(vi) exploit the VBF-like topology: two forward tagging jets, large $\Delta\eta_{jj}$ and high $m_{jj}$ with reduced central hadronic activity; while (iii) captures $s$-channel topologies where jets tend to be closer in phase space. This division follows standard LHC practices for VBF/di-jet tagging.
Observables (vii) are related to the kinematics of the decay $a\to W^+W^-\to e^\pm\mu^\mp\nu \bar{\nu}$ in the $e^\pm\mu^\mp+\met$ final state.

We implement the MVA using a BDT via the \textsc{TMVA} package integrated in \textsc{ROOT}, adopting the default BDT configuration for classification~\cite{TMVA:2007ngy}. 
This setup is used to maximize background rejection while retaining high signal efficiency.
\begin{figure}[h]
\centering
\includegraphics[width=0.48\linewidth]{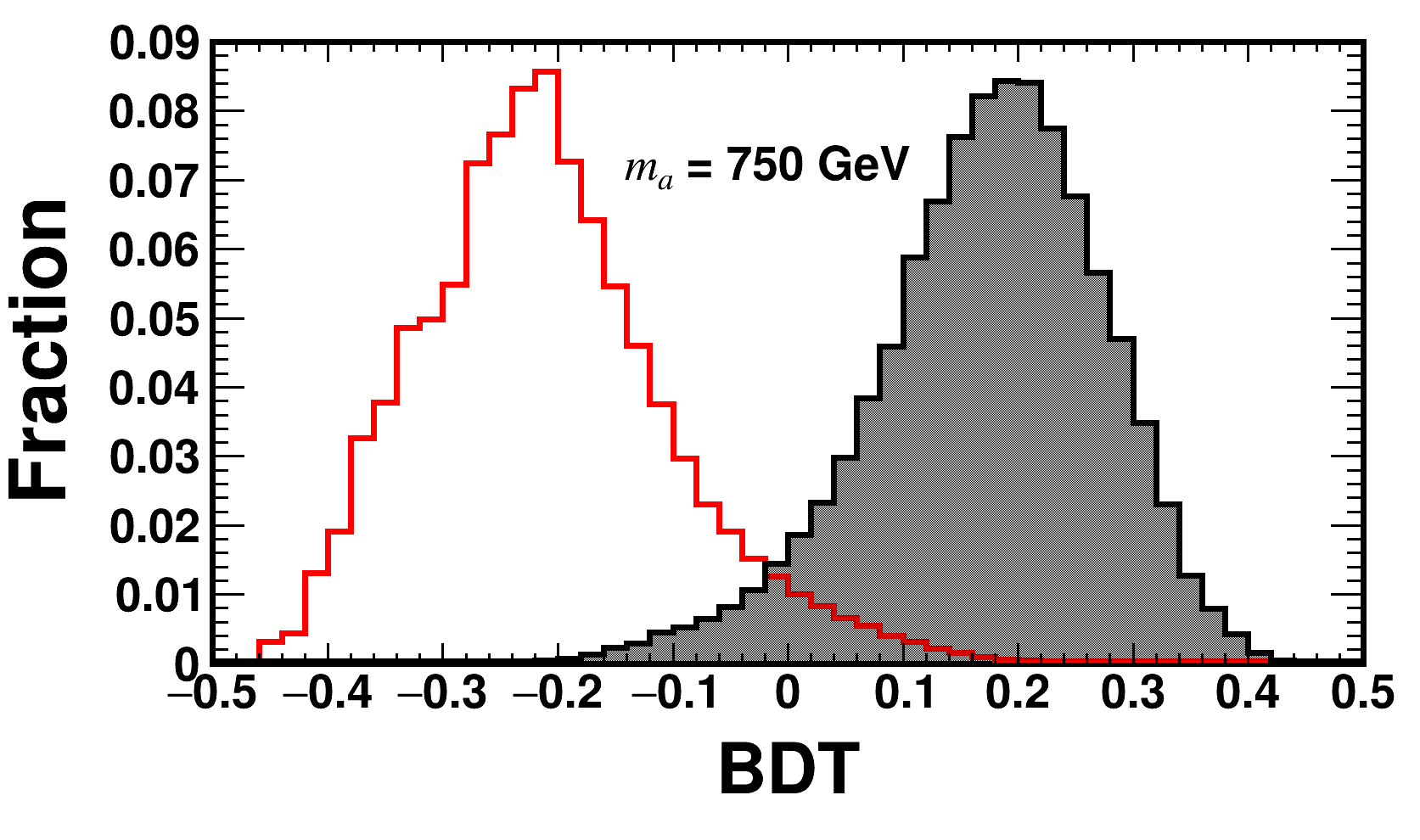}
\includegraphics[width=0.48\linewidth]{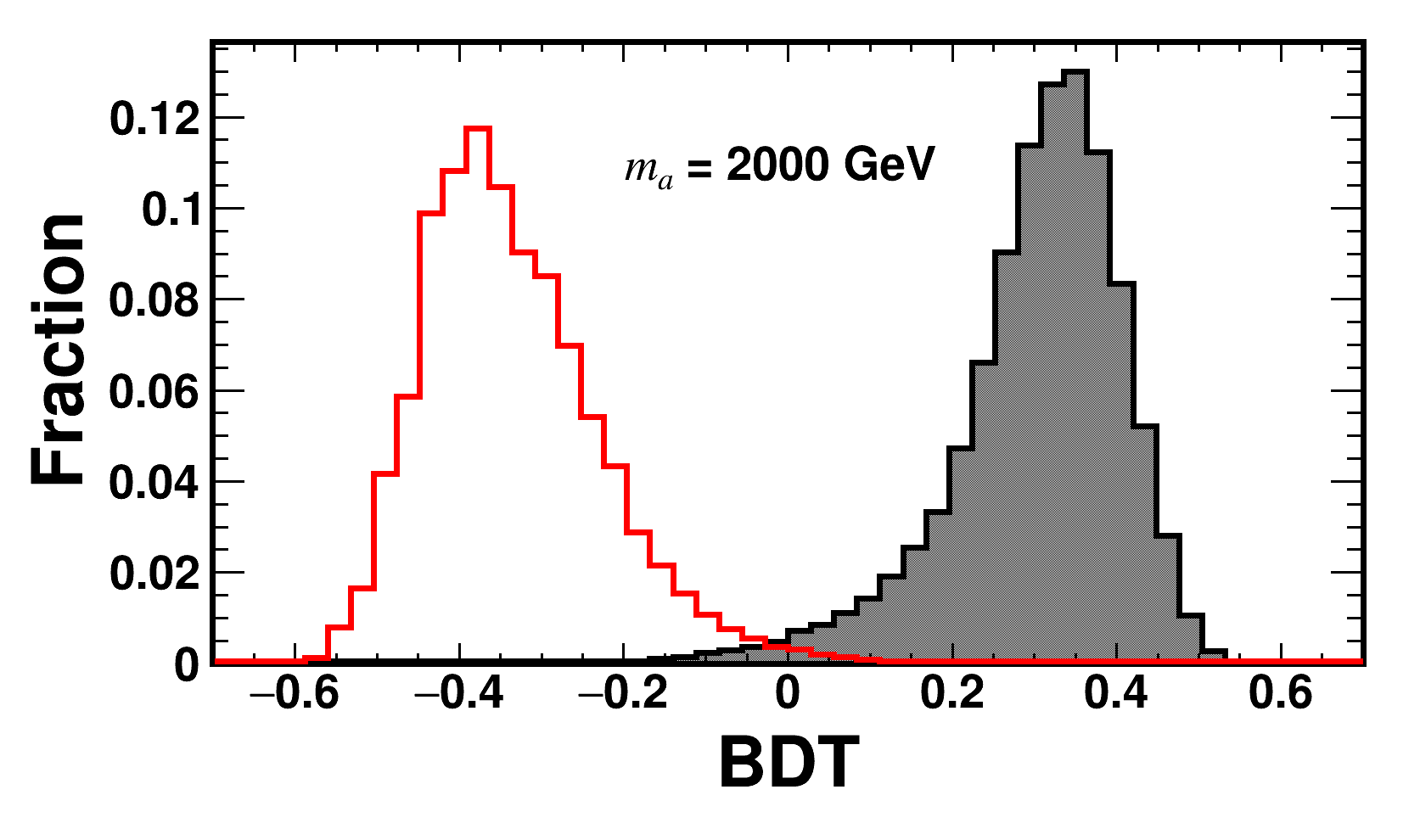}
\caption{BDT response distributions for signal (shaded) and total background (outline) at the HL-LHC with $\sqrt{s}=14~\mathrm{TeV}$, for $m_a=750~\mathrm{GeV}$ (left) and $m_a=2000~\mathrm{GeV}$ (right).}
\label{fig:bdt_response}
\end{figure}
Fig.~\ref{fig:bdt_response} shows the BDT responses for two benchmark masses, $m_a=750~\mathrm{GeV}$ and $m_a=2000~\mathrm{GeV}$. As $m_a$ increases, the underlying kinematics becomes more boosted and the separation between signal and background events improves, as seen by the shift of the signal (background) distribution toward higher (lower) BDT scores.

The BDT score threshold is optimized independently at each $m_a$ by maximizing the 
statistical significance,
\begin{equation}
\sigma_{\rm stat}=\sqrt{\,2\Big[\,(N_s+N_b)\,\ln\!\Big(1+\frac{N_s}{N_b}\Big)-N_s\,\Big]\,},
\label{eqn:Stat_Sgf}
\end{equation}
where $N_s$ and $N_b$ denote the expected signal and total background yields after preselection and the BDT cut~\cite{Cowan:2010js}. 

For completeness, Appendix~\ref{app:obs} displays representative input-observable distributions after preselection for benchmark $m_a=750~\mathrm{GeV}$, 
while Appendix~\ref{app:BDT} shows BDT responses across the full $m_a$ range for various $m_a$ cases. Appendix~\ref{app:BDTefficiency} reports the preselection and BDT efficiencies for signals and backgrounds at $\sqrt{s}=14~\mathrm{TeV}$. In that table, a dash (``$-$'') indicates that the surviving yield of a given background is negligible at $\mathcal{L}=3~\mathrm{ab}^{-1}$. Despite their large inclusive cross sections, the $W^\pm(\to\ell^\pm\nu)jj$ and $Z(\to\ell^+\ell^-)jj$ background events are almost completely removed by the opposite-sign $e\mu$ requirement and the multivariate selection, so they are omitted from the appendix table for brevity.

\section{Results}
\label{sec:results}

Building on the strategy above, we present the projected discovery sensitivities at the HL-LHC with $\sqrt{s}=14~\mathrm{TeV}$ and $\mathcal{L}=3~\mathrm{ab}^{-1}$. We show both the discovery reach on the coupling $g_{aWW}$ and the corresponding model-independent reach on $\sigma(pp\to jj\,a)\times \mathrm{Br}(a\to W^+W^-)$ across $m_a\in[170,4000]~\mathrm{GeV}$.
Statistical significances are evaluated with Eq.~(\ref{eqn:Stat_Sgf}). All results assume the \emph{photophobic} limit $g_{a\gamma\gamma}=0$ as defined in Ref.~\cite{Craig:2018kne}; the BDT score threshold is optimized independently at each mass point (see Sec.~\ref{subsec:multi}). Unless stated otherwise, the quoted sensitivities are statistical only; the impact of systematics can be estimated in reinterpretations by introducing background-normalization nuisances in a profile-likelihood fit following Ref.~\cite{Cowan:2010js}. 

\begin{figure}[h]
\centering
\includegraphics[width=\linewidth]{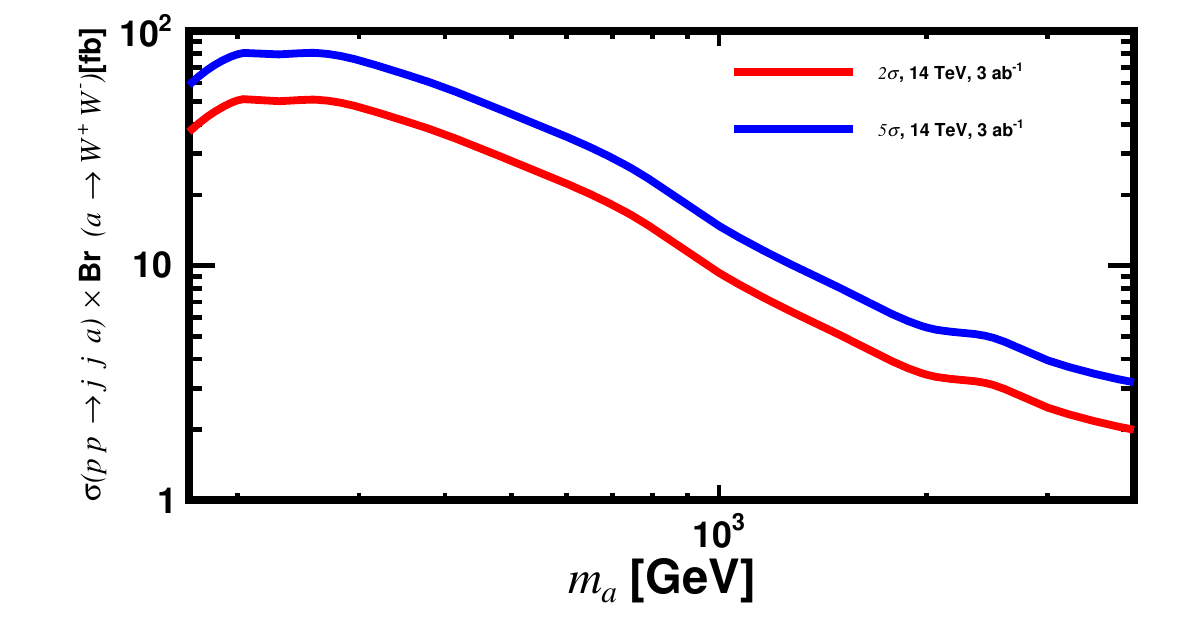}
\caption{Projected $2\sigma$ and $5\sigma$ discovery sensitivities on the model-independent fiducial production cross section $\sigma(pp\to jj\,a)\times \mathrm{Br}(a\to W^+W^-)$ versus $m_a$ at the HL-LHC ($\sqrt{s}=14~\mathrm{TeV}$, $\mathcal{L}=3~\mathrm{ab}^{-1}$). 
Assumes $g_{a\gamma\gamma}=0$ and per-mass BDT optimization.
Red and blue curves correspond to $2\sigma$ and $5\sigma$, respectively.}
\label{fig:xsec_sensitivity}
\end{figure}

Fig.~\ref{fig:xsec_sensitivity} reports the corresponding $2\sigma$ and $5\sigma$ sensitivities on the model-independent fiducial production cross section $\sigma(pp\to jj\,a)\times \mathrm{Br}(a\to W^+W^-)$, facilitating reinterpretations in alternative theory models. Near the threshold ($m_a\sim170$–$200~\mathrm{GeV}$), the sensitivity varies quickly as the $W^+W^-$ mode turns on and the event kinematics evolve; while in larger $m_a$ region, the signal–background separation improves (cf.\ Sec.~\ref{subsec:multi}), which offsets the falling production rate and allows competitiveness up to multi-TeV masses.

\begin{figure}[h]
\centering
\includegraphics[width=\linewidth]{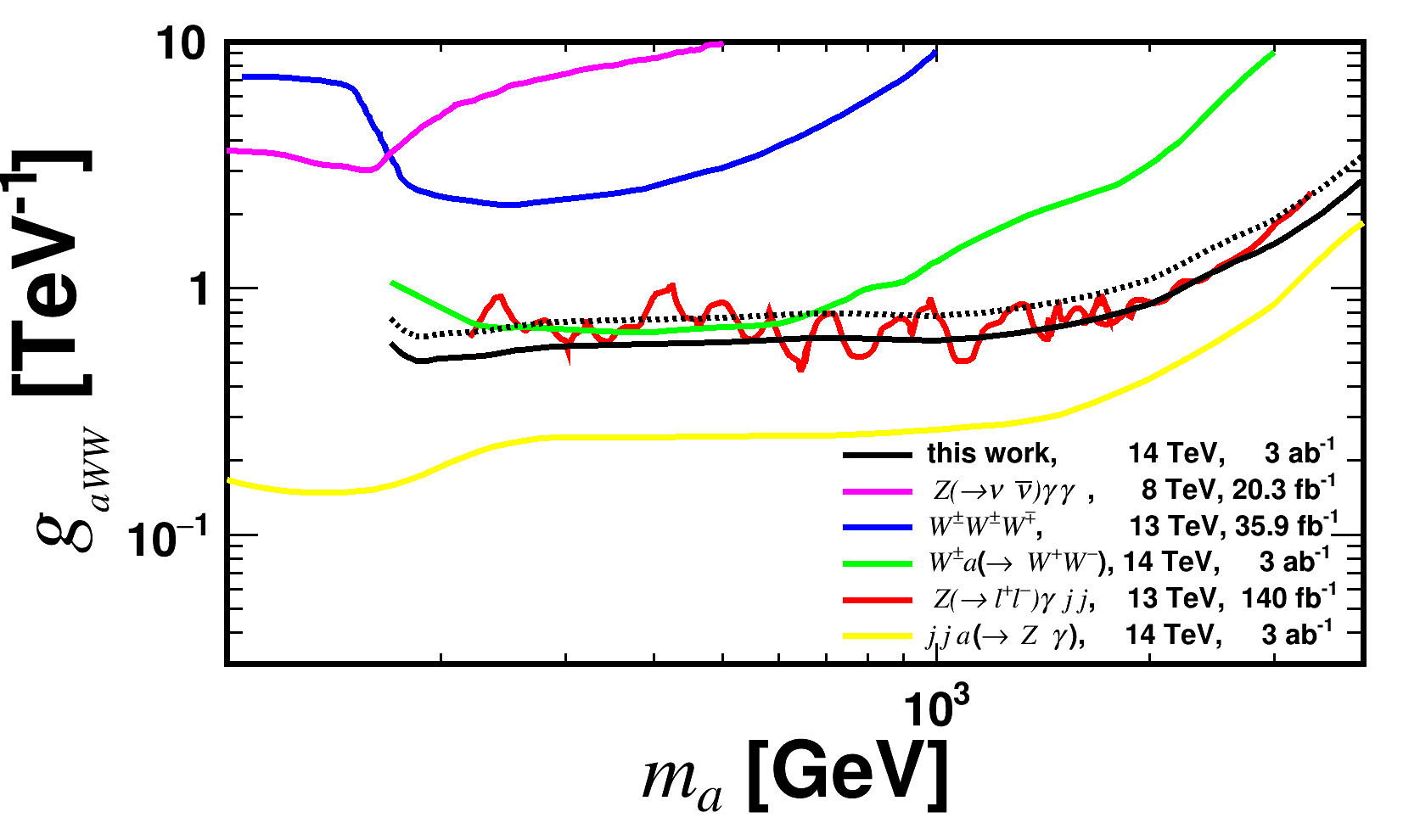}
\caption{
Projected discovery reach on $g_{aWW}$ for photophobic ALPs as a function of $m_a$ (170–4000~GeV) at the HL-LHC ($\sqrt{s}=14~\mathrm{TeV}$, $\mathcal{L}=3~\mathrm{ab}^{-1}$). Solid (dashed) black curves denote $2\sigma$ ($5\sigma$) significances. The projection assumes $g_{a\gamma\gamma}=0$ and uses a per–mass-point BDT optimization; significances are statistical only. For comparison, we overlay $95\%$~CL limits recast from triboson 
$pp \to Z (\to \nu\bar{\nu}) \gamma \gamma$
at 8~TeV with 20.3~fb$^{-1}$~\cite{Craig:2018kne}, and from Run-2 analyses at 13~TeV: $pp\!\to\!W^\pm W^\pm W^\mp$ with 35.9~fb$^{-1}$ and $pp\!\to\!Z(\to\ell^+\ell^-)\gamma jj$ with 140~fb$^{-1}$~\cite{Aiko:2024xiv}. Also shown are HL-LHC ($14~\mathrm{TeV}$, $3~\mathrm{ab}^{-1}$) $2\sigma$ projections for $pp\!\to\!W^\pm a(\to W^+W^-)$~\cite{Mao:2024kgx} and $pp\!\to\!jj\,a(\to Z\gamma)$~\cite{Ding:2024djo}.
}
\label{fig:conclusion}
\end{figure}

Fig.~\ref{fig:conclusion} summarizes the $2\sigma$ and $5\sigma$ discovery reaches on $g_{aWW}$ for heavy photophobic ALPs. 
At $m_a=170~\mathrm{GeV}$, the $2\sigma\,(5\sigma)$ discovery threshold is $g_{aWW}=0.60\,(0.76)~\mathrm{TeV}^{-1}$. As the $a\to W^+W^-$ channel opens and production cross section of $pp\to jj\,a(\to W^\pm W^\mp)$ grows rapidly between $170$ and $200~\mathrm{GeV}$ (cf. Fig.~\ref{fig:crosssection}), the required coupling decreases; over $260\!\le\! m_a\!\le\!1500~\mathrm{GeV}$ it remains near $0.61\,(0.76)~\mathrm{TeV}^{-1}$. At very high masses the reach weakens due to falling PDFs and reduced acceptance, with the $2\sigma\,(5\sigma)$ discovery threshold $g_{aWW}=2.75\,(3.48)~\mathrm{TeV}^{-1}$ at $m_a=4~\mathrm{TeV}$.

Fig.~\ref{fig:conclusion} also displays previous constraints/projections for context. 
The magenta curve corresponds to the $95\%$~C.L. limit obtained by reinterpreting the Run-1 triboson channel $pp\!\to\!Z(\to\nu\bar\nu)\gamma\gamma$ at $\sqrt{s}=8$~TeV with $20.3~\mathrm{fb}^{-1}$~\cite{Craig:2018kne}. 
The blue and red curves show Run-2 recasts at $\sqrt{s}=13$~TeV: $pp\!\to\!W^\pm W^\pm W^\mp$ with $35.9~\mathrm{fb}^{-1}$ and $pp\!\to\!Z(\to\ell^+\ell^-)\gamma jj$ with $140~\mathrm{fb}^{-1}$, respectively, as reported in Ref.~\cite{Aiko:2024xiv}. 
The green curve is an HL-LHC $2\sigma$ projection for the tri-$W$ topology $pp\!\to\!W^\pm a(\to W^+W^-)$ using a same-sign dimuon plus hadronic-$W$ selection and an MVA~\cite{Mao:2024kgx}. 
The yellow curve shows the HL-LHC $2\sigma$ projection for $pp\!\to\!jj\,a(\to Z\gamma)$, which exploits a clean $Z(\to\ell^+\ell^-)$+$\gamma$ resonance with both the VBF-like and s-channel productions and detector-level MVA~\cite{Ding:2024djo}.

Relative to the tri-$W$ strategy (green), our dedicated $jj\, a(\to W^+ W^-)$ analysis achieves a stronger reach (lower $g_{aWW}$) over all mass ranges, with a particularly visible improvement at high mass region ($m_a \gtrsim$ 700 GeV), where the VBF-like kinematics sharpen signal–background separation. 
When compared to the $jj\,a(\to Z\gamma)$ projection (yellow), our reach is weaker across the mass range. 
The reason is clear: the $jj\,a(\to Z\gamma)$ final state benefits from a narrow mass peak, excellent photon/lepton resolution, and lower irreducible backgrounds~\cite{Ding:2024djo}; by contrast, although the branching ratio of $a \to W^+ W^-$ is larger than that of $a \to Z \gamma$, the dileptonic-$WW$ channel involves missing energy, larger SM backgrounds ($WW$, $t\bar t$), and no fully reconstructible resonance, which intrinsically reduces sensitivity.
Despite this, the $jj\, a(\to W^+ W^-)$ topology remains important and complementary. 
First, it \emph{directly} probes the $aWW$ vertex via $a\!\to\!W^+W^-$, while $a \to Z\gamma$ tests the $aZ\gamma$ interaction; a joint observation/non-observation across channels would enable a nontrivial check of the photophobic relations $g_{aZ\gamma}=t_\theta\,g_{aWW}$ and $g_{aZZ}=(1-t_\theta^2)\,g_{aWW}$ and thus of the underlying electroweak structure~\cite{Craig:2018kne,Aiko:2024xiv}. 
Second, systematics and backgrounds are largely orthogonal between $jj\, a(\to W^+ W^-)$ and $jj\,a(\to Z\gamma)$, providing robustness against analysis-specific effects and a richer basis for global fits. 
Third, we provide model-independent discovery thresholds for $\sigma(pp\!\to\!jj\,a)\times\mathrm{Br}(a\!\to\!W^+W^-)$, which enable straightforward reinterpretation for other neutral resonances decaying to $WW$ (including scenarios where $aZ\gamma$ is suppressed but $aWW$ remains sizable)~\cite{Mao:2024kgx}. 
Taken together, these considerations warrant a dedicated $jj\, a(\to W^+ W^-)$ search in parallel with $Z\gamma$ analyses.

\section{Conclusions}
\label{sec:conc}

We have assessed the discovery potential for heavy \emph{photophobic} ALPs at the HL-LHC ($\sqrt{s}=14~\mathrm{TeV}$), where the ALP--diphoton coupling vanishes, $g_{a\gamma\gamma}=0$, and mainly interactions with electroweak gauge bosons~\cite{Craig:2018kne}. 
In this limit, the ALP is produced in association with two jets, $pp\to jj\,a$, via either $s$-channel electroweak-boson exchange or VBF-like topologies; for $m_a\gtrsim 2\,m_W$, the decay channel $a\to W^+W^-$ dominates.
We target the opposite-sign, different-flavour dilepton final state $e^\pm\mu^\mp+jj+\met$, which strongly suppresses Drell--Yan contamination and multijet backgrounds, and we adopt the HL-LHC baseline of $\sqrt{s}=14~\mathrm{TeV}$ and $\mathcal{L}=3~\mathrm{ab}^{-1}$~\cite{CERN-2019-007}. 

Events are generated with MadGraph5\_aMC@NLO, showered and hadronised with \textsc{PYTHIA}~8.3, and passed through a Delphes~3 fast detector simulation~\cite{Alwall:2014hca,Bierlich:2022pfr,deFavereau:2013fsa}. The analysis applies a physics-motivated preselection for the $e\mu+jj+\met$ topology and then a multivariate classification based on a BDT implemented in \textsc{TMVA}~\cite{TMVA:2007ngy}. The BDT input features capture both VBF-like jet–gap kinematics (large $m_{jj}$ and $\Delta\eta_{jj}$, reduced central activity) and $s$-channel configurations (small $\Delta R_{jj}$ and reconstructed $m_{jj} \sim m_V$), and the BDT score threshold is optimised independently at each $m_a$. 

Our results provide the projected discovery reaches at $2\sigma$ and $5\sigma$ statistical significances on the ALP--$W$ coupling $g_{aWW}$ and, in a model-independent form, on $\sigma(pp\to jj\,a)\times\mathrm{Br}(a\to W^+W^-)$, over the mass range $m_a\in[170,4000]~\mathrm{GeV}$.
Specifically, 
for $260\le m_a\le 1500~\mathrm{GeV}$, the sensitivity remains approximately constant around $0.61\,(0.76)~\mathrm{TeV}^{-1}$ at the $2\sigma\,(5\sigma)$ level.
We also report discovery sensitivities on the fiducial production cross sections $\sigma(pp\to jj\,a)\times\mathrm{Br}(a\to W^+W^-)$ across $m_a\in[170,4000]~\mathrm{GeV}$, enabling straightforward reinterpretations in alternative theory models. 

For orientation, we compare our $p p \to j j a\,(\to W^+ W^-)$ projection with recent HL-LHC studies and Run-1/2 recasts: the tri-$W$ strategy $pp\!\to\!W^\pm a(\to W^+W^-)$~\cite{Mao:2024kgx}, the $jj\,a(\to Z\gamma)$ channel~\cite{Ding:2024djo}, and related recasts~\cite{Craig:2018kne,Aiko:2024xiv}. In short, our reach improves on the tri-$W$ strategy (notably at high $m_a$) but is weaker than the clean $Z\gamma$ resonance channel; see Fig.~\ref{fig:conclusion}. The $WWjj$ topology remains complementary: it directly probes $aWW$, carries largely orthogonal systematics to $Z\gamma$, and provides model-independent thresholds in $\sigma(pp\!\to\!jj\,a)\times\mathrm{Br}(a\!\to\!W^+W^-)$; full details and a quantitative comparison are deferred to Sec.~\ref{sec:results}.

In summary, the $e\mu+jj+\met$ topology at the HL-LHC offers competitive sensitivity to heavy photophobic ALPs, with robust performance in the multi-TeV regime thanks to improved signal–background separation at large $m_a$. The model-independent $\sigma\times\mathrm{Br}$ projections are intended to facilitate recasts beyond the specific benchmark ALP model studied in this work.

\appendix

\newpage
\section{Representative observable distributions}
\label{app:obs}

\begin{figure}[htbp] 
\centering
%\addtocounter{figure}{-1}
\subfigure{
\includegraphics[width=7.3cm,height=4.2cm]{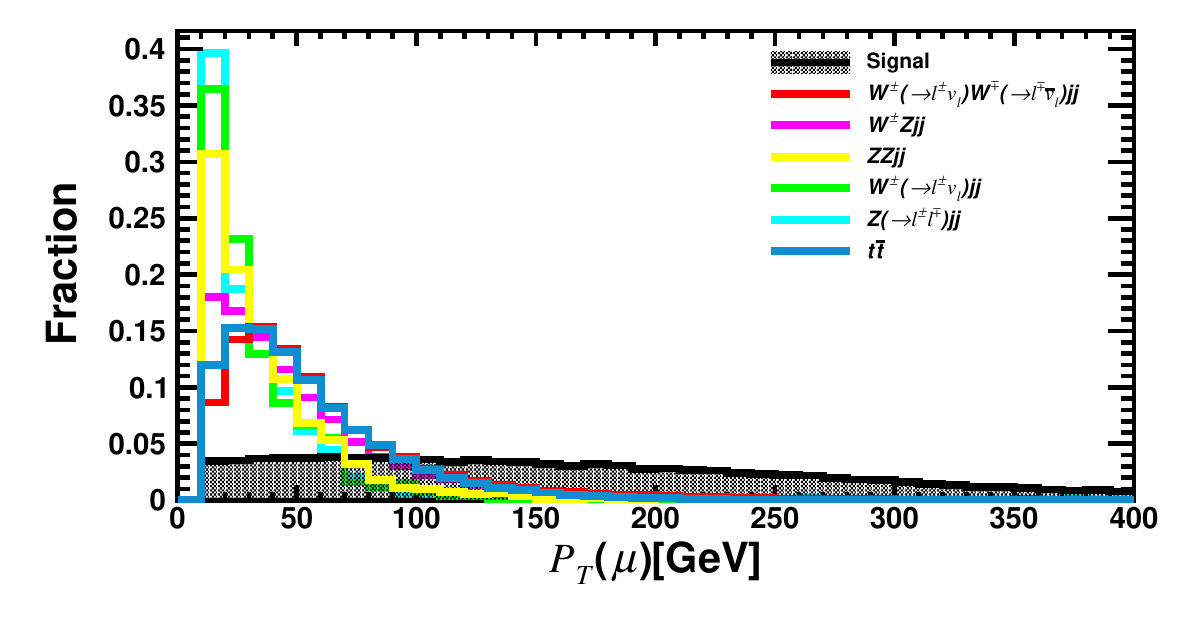}	
\includegraphics[width=7.3cm,height=4.2cm]{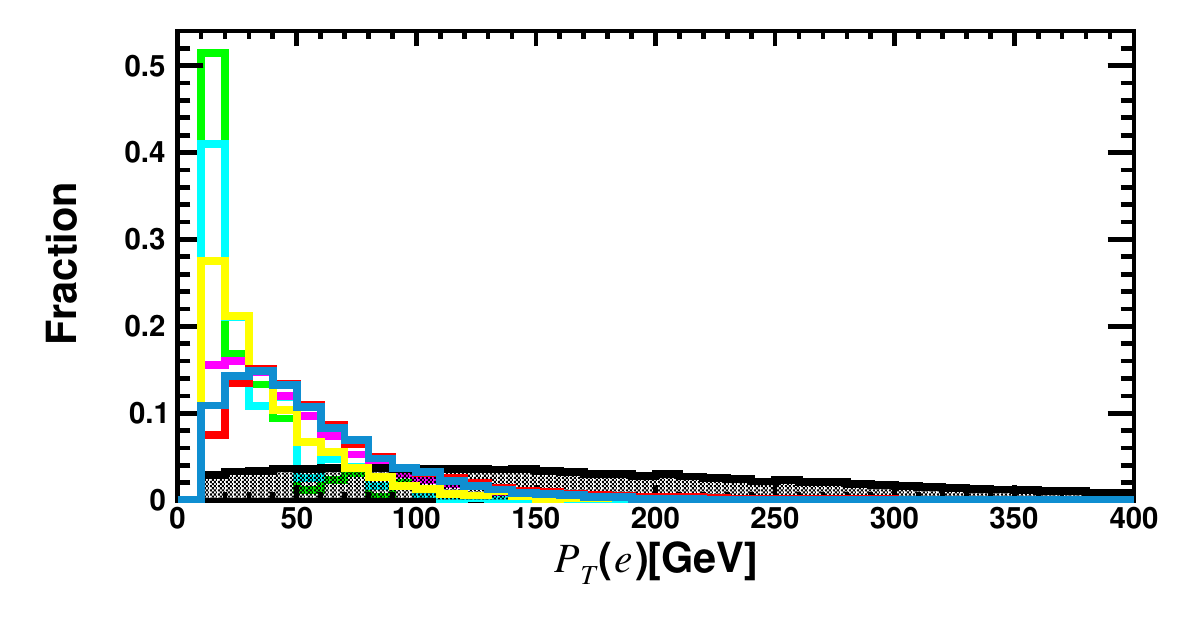}
}
\end{figure}
\vspace{-1.0cm}
\begin{figure}[htbp] 
\centering
%\addtocounter{figure}{1}
\subfigure{
\includegraphics[width=7.3cm,height=4.2cm]{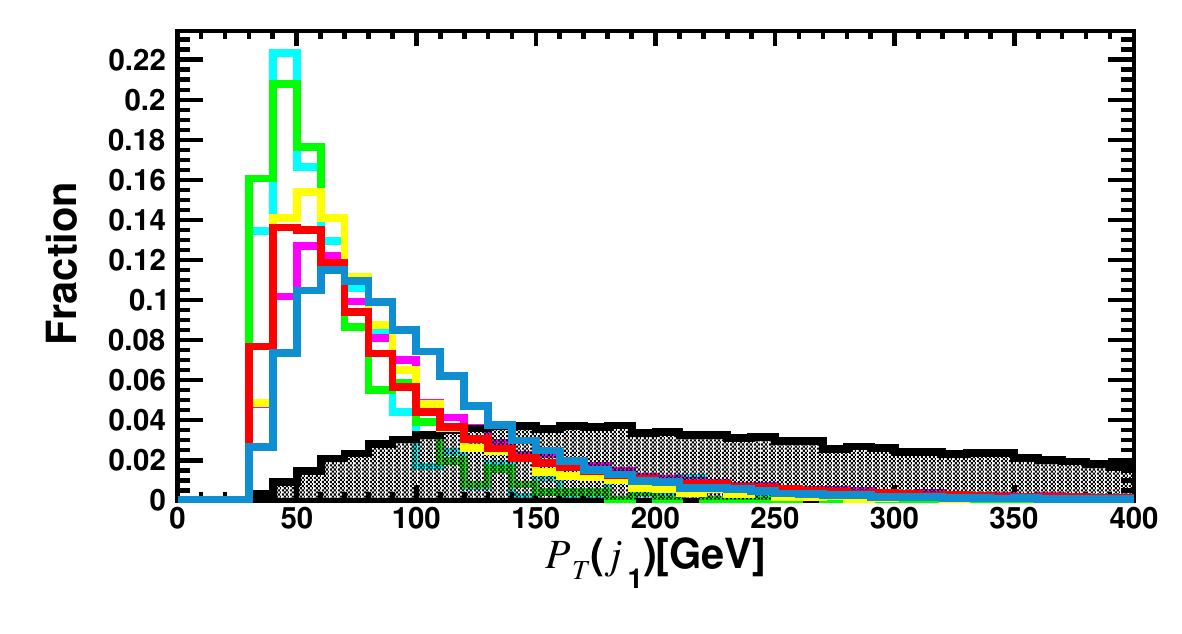}
\includegraphics[width=7.3cm,height=4.2cm]{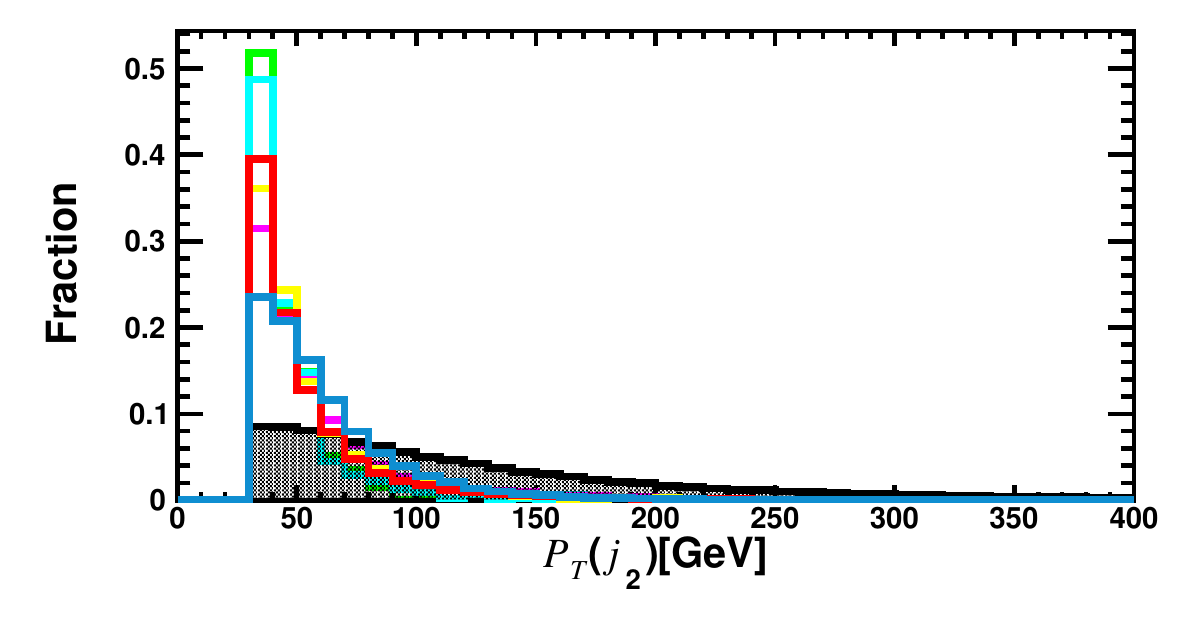}
}
\end{figure}
\vspace{-1.0cm}
\begin{figure}[htbp] 
\centering
%\addtocounter{figure}{-1}
\subfigure{
\includegraphics[width=7.3cm,height=4.2cm]{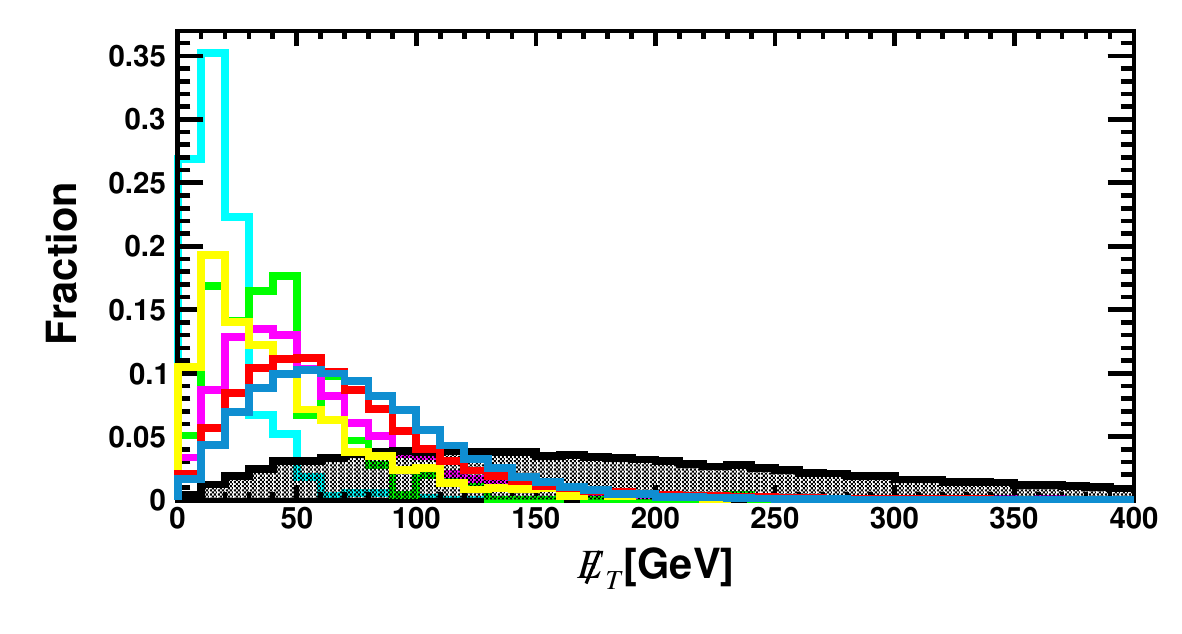}
\includegraphics[width=7.3cm,height=4.2cm]{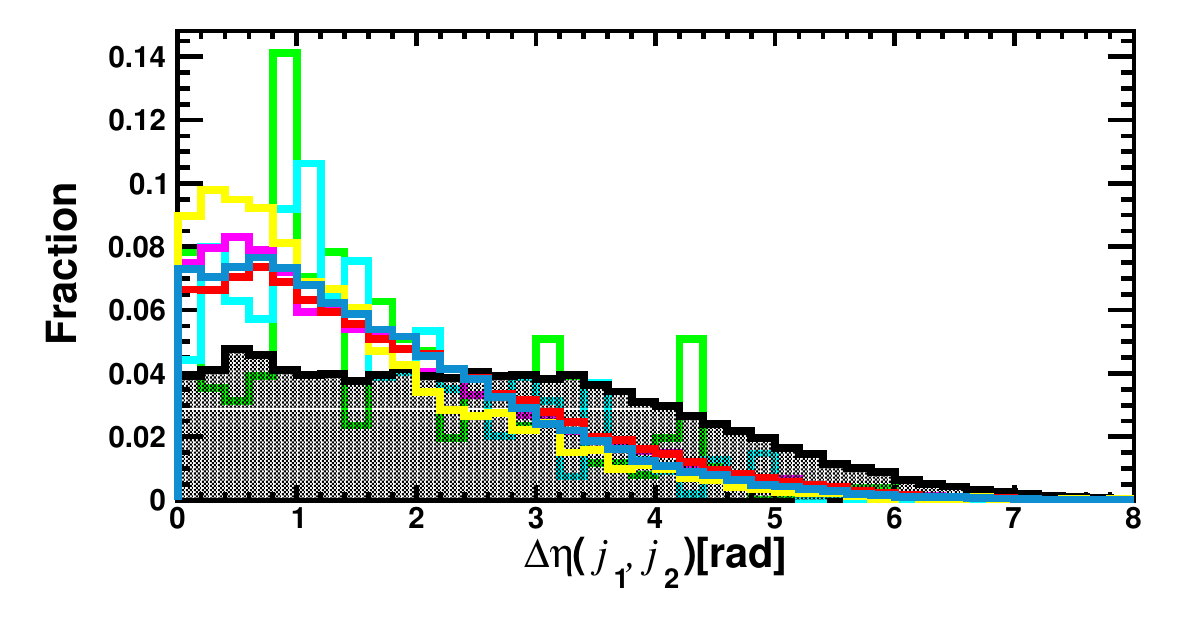}
}
\end{figure}
\vspace{-1.0cm}
\begin{figure}[htbp] 
\centering
%\addtocounter{figure}{-1}
\subfigure{
\includegraphics[width=7.3cm,height=4.2cm]{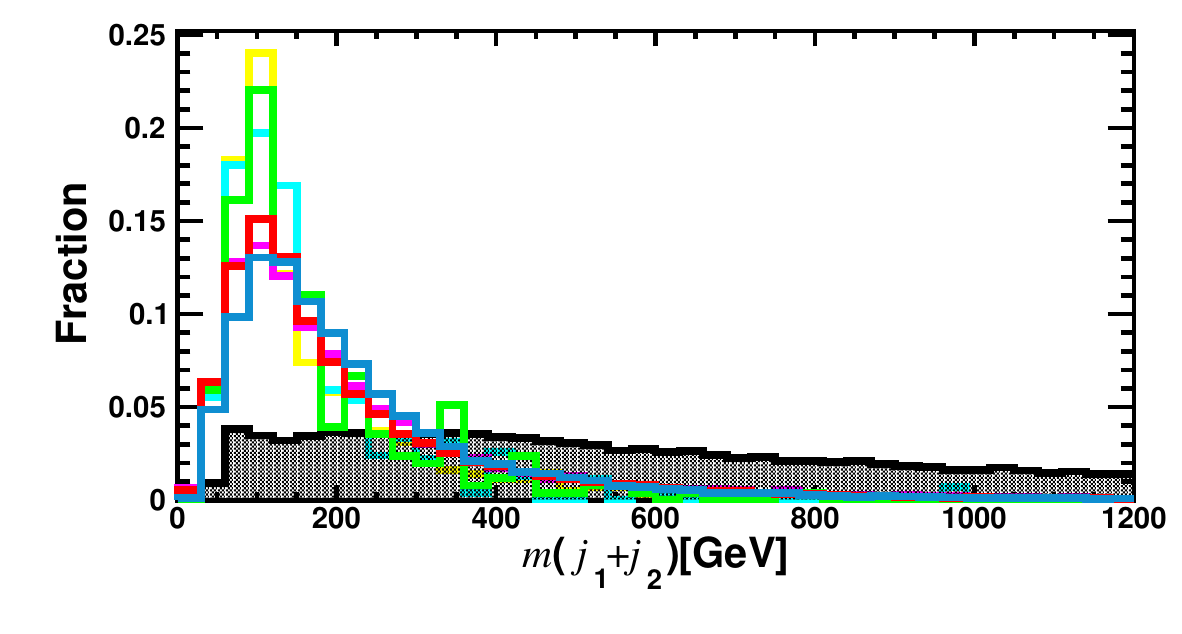}
\includegraphics[width=7.3cm,height=4.2cm]{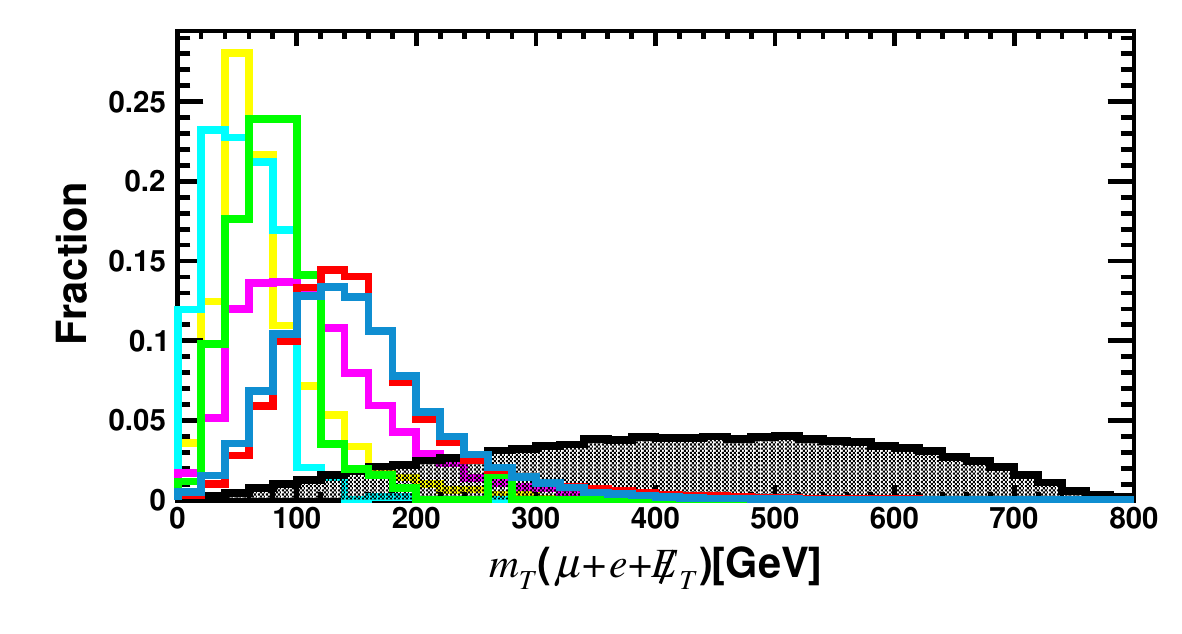}
}
\caption{
Representative observable distributions for the signal (black, dashed) and the six background processes at the HL-LHC ($\sqrt{s}=14~\mathrm{TeV}$), assuming $m_a=750~\mathrm{GeV}$ and after the preselection of Section~\ref{subsec:presel}. 
}
\label{fig:obs14TeV}
\end{figure}

\newpage
\section{BDT response distributions}
\label{app:BDT}

\begin{figure}[htbp]
\centering
\subfigure
{
\includegraphics[width=7.3cm,height=4.2cm]{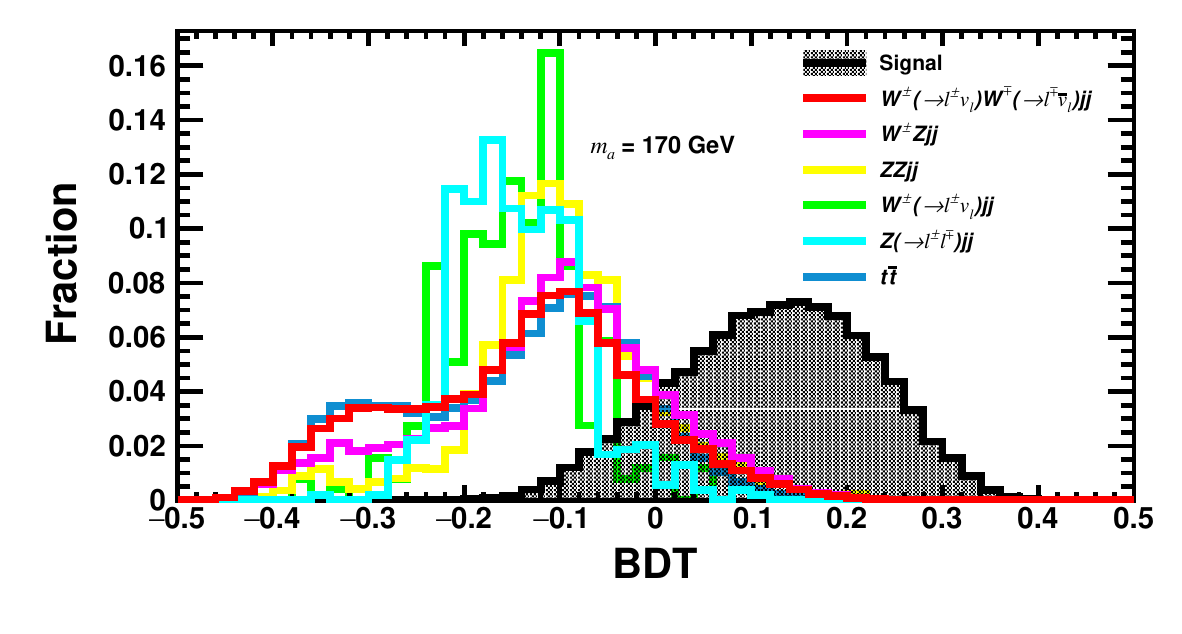}
\includegraphics[width=7.3cm,height=4.2cm]{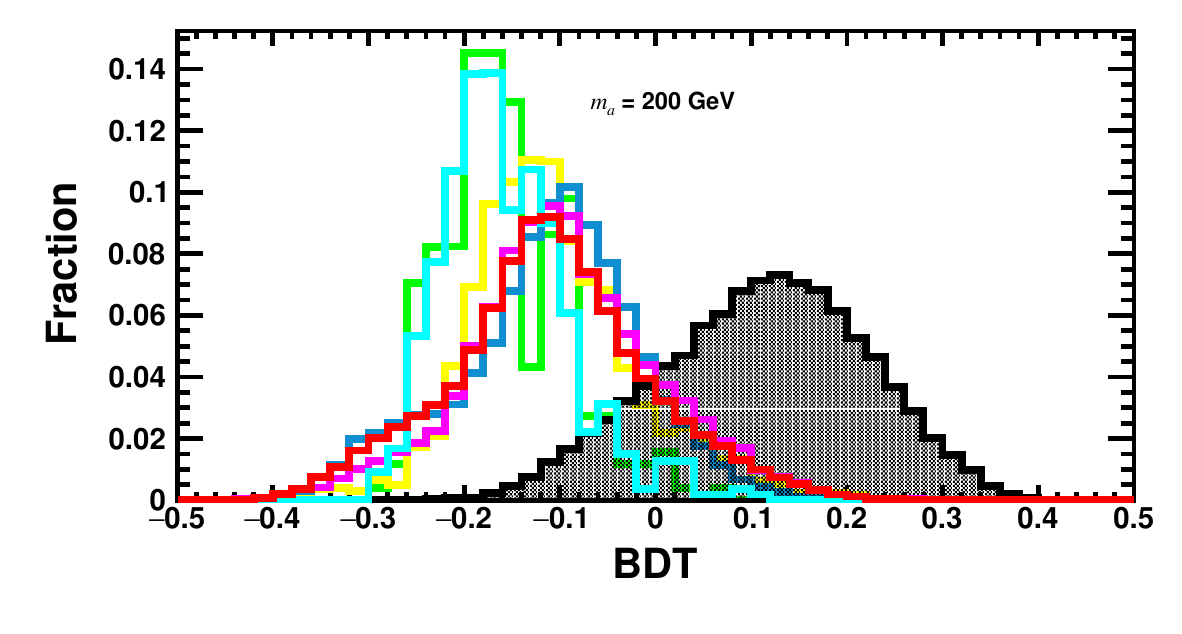}
}
\end{figure}
%\addtocounter{figure}{-1}
%
\vspace{-1.0cm}
\begin{figure}[htbp]
\centering
\subfigure
{
\includegraphics[width=7.3cm,height=4.2cm]{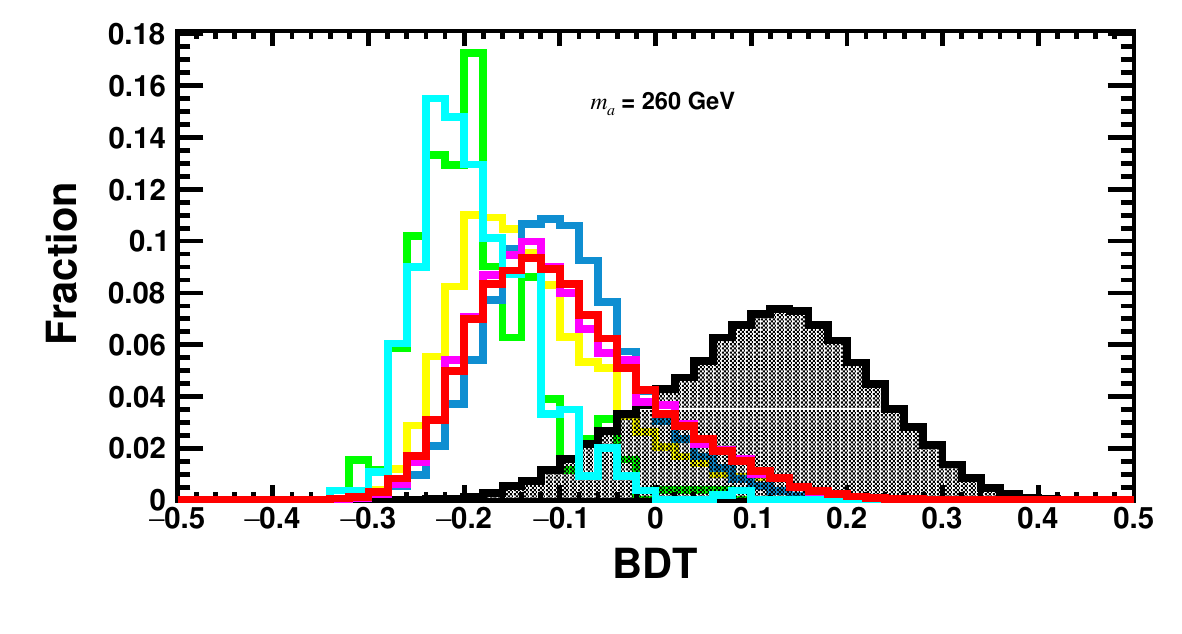}
\includegraphics[width=7.3cm,height=4.2cm]{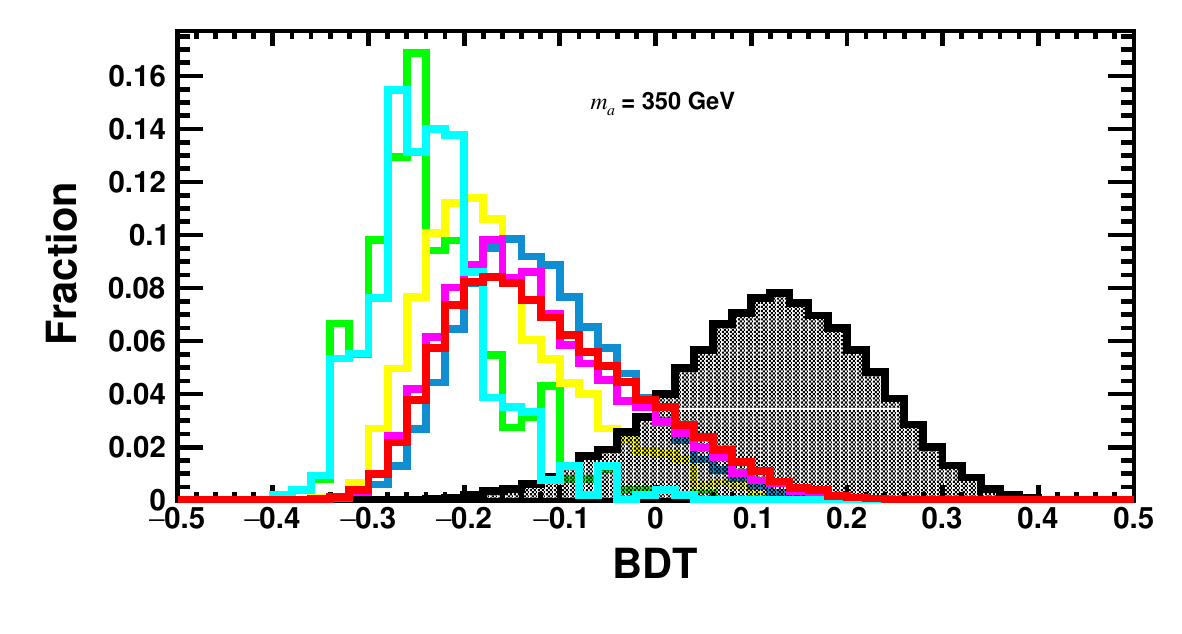}
}
\end{figure}
%\addtocounter{figure}{-1}
%
\vspace{-1.0cm}
\begin{figure}[htbp] 
\centering
%\addtocounter{figure}{1}
\subfigure{
\includegraphics[width=7.3cm,height=4.2cm]{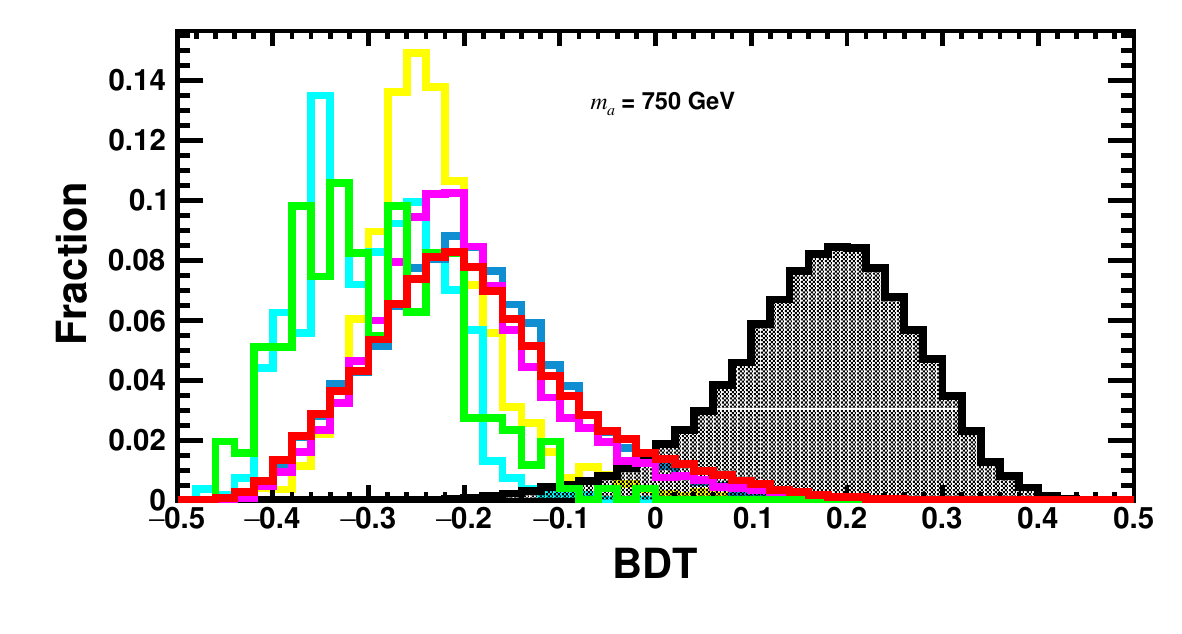}
\includegraphics[width=7.3cm,height=4.2cm]{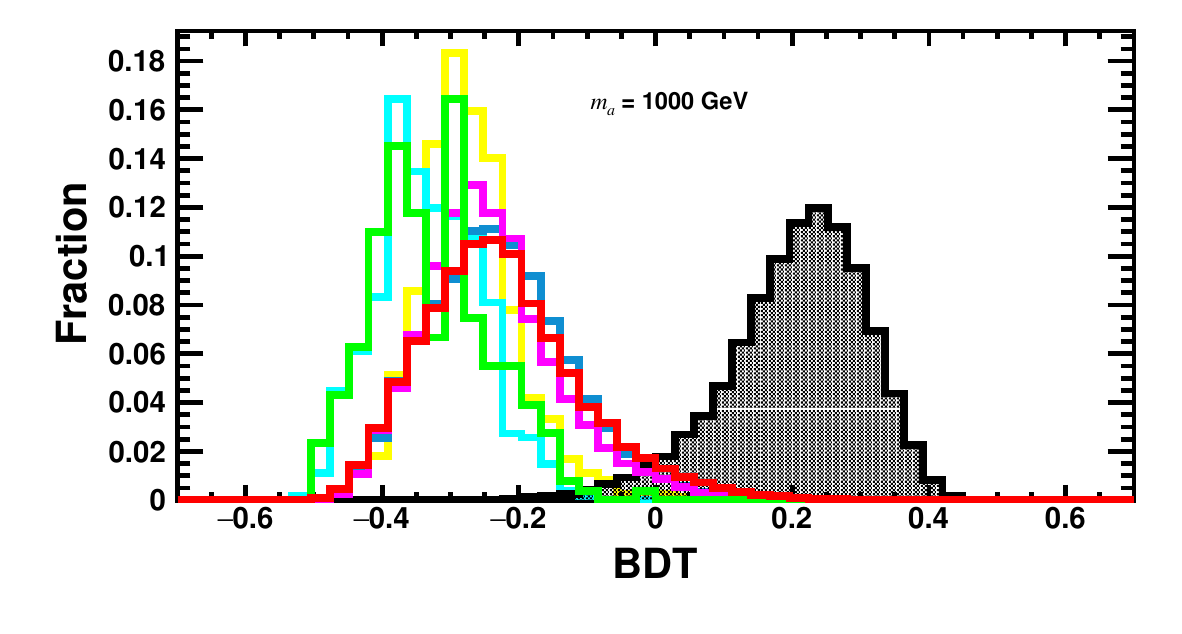}
}
\end{figure}
\vspace{-1.0cm}
\begin{figure}[htbp] 
\centering
%\addtocounter{figure}{1}
\subfigure{
\includegraphics[width=7.3cm,height=4.2cm]{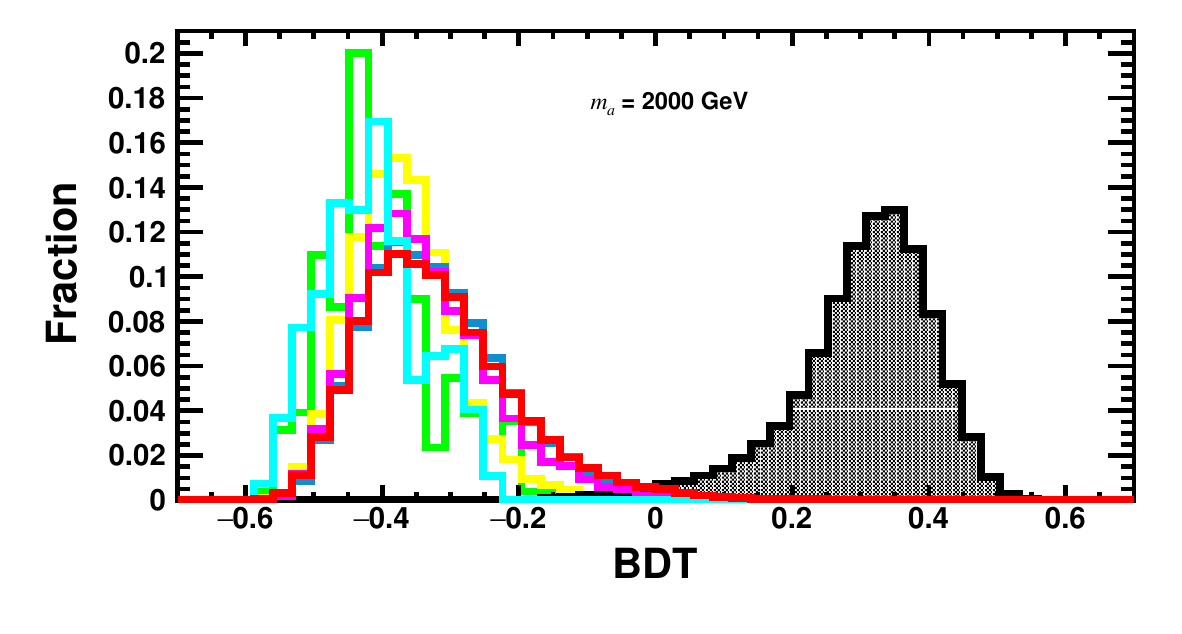}
\includegraphics[width=7.3cm,height=4.2cm]{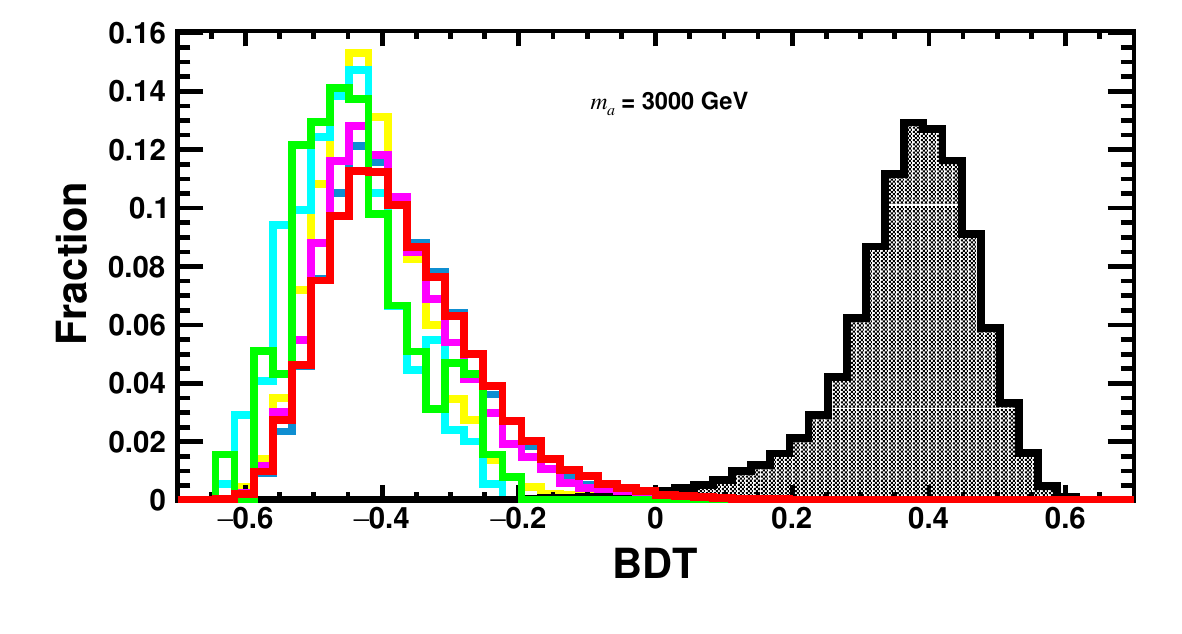}
}
\caption{BDT response distributions for the signal (shaded) and the six background processes at the HL-LHC ($\sqrt{s}=14~\mathrm{TeV}$), for representative ALP masses after the preselection of Section~\ref{subsec:presel}.
}
\label{fig:BDT14TeV}
\end{figure}

\newpage
\section{The selection efficiency table}
\label{app:BDTefficiency}

\begin{table}[h]
\centering
\small
\setlength{\tabcolsep}{6pt}
\caption{BDT working points and post-BDT selection efficiencies at the HL-LHC ($\sqrt{s}=14~\mathrm{TeV}$) for various $m_a$. The second column gives the lower BDT-score threshold (cut value) used at each mass point; entries report the fraction of events surviving the BDT \emph{after the preselection} of Section~\ref{subsec:presel}. A dash (``$-$'') indicates a negligible yield at $\mathcal{L}=3~\mathrm{ab}^{-1}$.}
\label{tab:BDT}
\begin{tabular}{lcccccc}
\hline\hline
$m_a$ [GeV] & BDT cut & Signal & $W^+ (\to \ell^+ \nu) \, W^-(\to \ell^- \bar{\nu}) jj$ & $WZjj$ & $ZZjj$ & $t\bar t$ \\
\hline
170  & 0.165 & $3.62\times10^{-1}$ & $4.91\times10^{-3}$ & $4.74\times10^{-3}$ & $3.85\times10^{-3}$ & $2.77\times10^{-3}$ \\
185  & 0.144 & $4.06\times10^{-1}$ & $9.07\times10^{-3}$ & $8.22\times10^{-3}$ & $6.93\times10^{-3}$ & $6.58\times10^{-3}$ \\
200  & 0.141 & $4.14\times10^{-1}$ & $1.24\times10^{-2}$ & $1.04\times10^{-2}$ & $5.39\times10^{-3}$ & $8.57\times10^{-3}$ \\
230  & 0.146 & $3.91\times10^{-1}$ & $1.13\times10^{-2}$ & $9.96\times10^{-3}$ & $4.62\times10^{-3}$ & $9.35\times10^{-3}$ \\
260  & 0.144 & $4.02\times10^{-1}$ & $1.28\times10^{-2}$ & $8.99\times10^{-3}$ & $4.24\times10^{-3}$ & $9.52\times10^{-3}$ \\
350  & 0.152 & $3.85\times10^{-1}$ & $9.02\times10^{-3}$ & $4.54\times10^{-3}$ & $7.70\times10^{-4}$ & $7.45\times10^{-3}$ \\
500  & 0.166 & $4.03\times10^{-1}$ & $5.33\times10^{-3}$ & $2.22\times10^{-3}$ & $-$                & $3.12\times10^{-3}$ \\
750  & 0.176 & $5.19\times10^{-1}$ & $3.48\times10^{-3}$ & $1.16\times10^{-3}$ & $-$                & $1.99\times10^{-3}$ \\
1000 & 0.204 & $5.56\times10^{-1}$ & $1.43\times10^{-3}$ & $2.90\times10^{-4}$ & $-$                & $3.46\times10^{-4}$ \\
1500 & 0.231 & $6.92\times10^{-1}$ & $7.41\times10^{-4}$ & $1.93\times10^{-4}$ & $-$                & $1.73\times10^{-4}$ \\
2000 & 0.243 & $7.75\times10^{-1}$ & $4.48\times10^{-4}$ & $9.67\times10^{-5}$ & $-$                & $8.66\times10^{-5}$ \\
2500 & 0.230 & $8.74\times10^{-1}$ & $4.66\times10^{-4}$ & $1.93\times10^{-4}$ & $-$                & $8.66\times10^{-5}$ \\
3000 & 0.228 & $9.07\times10^{-1}$ & $2.93\times10^{-4}$ & $9.67\times10^{-5}$ & $-$                & $8.66\times10^{-5}$ \\
4000 & 0.235 & $9.43\times10^{-1}$ & $2.07\times10^{-4}$ & $9.67\times10^{-5}$ & $-$                & $-$                \\
\hline\hline
\end{tabular}
\\[0.35em]
\begin{minipage}{0.96\linewidth}
\footnotesize \emph{Note.} The inclusive $Wjj$ and $Zjj$ backgrounds are omitted from this table: after the opposite-sign $e\mu$ preselection and the per-mass BDT optimisation (Section~\ref{subsec:multi}), their expected yields are negligible at $\mathcal{L}=3~\mathrm{ab}^{-1}$; see also Tables~\ref{tab:preselection14TeV} and \ref{tab:presel}.
\end{minipage}
\end{table}

\acknowledgments

We thank Bin Diao, Zilong Ding, Yiheng Xiong and Ye Lu for helpful discussions. 
J.F. and K.W. are supported by the National Natural Science Foundation of China under grant no. 11905162, the Excellent Young Talents Program of the Wuhan University of Technology under grant no. 40122102, and the research program of the Wuhan University of Technology under grants no. 2020IB024 and 104972025KFYjc0101. 
Y.N.M. is supported by the National Natural Science Foundation of China under grant no. 12205227. 
The simulation and analysis work of this article was completed with the computational cluster provided by the Theoretical Physics Group at the Department of Physics, School of Physics and Mechanics, Wuhan University of Technology.

\paragraph{Note added.}
Following standard practice in high energy physics, authors are listed in strict alphabetical order by surname. 
This ordering should not be interpreted as indicating any ranking of contribution, seniority, or leadership.
All authors contributed equally to this work.

% Bibliography

%% [A] Recommended: using JHEP.bst file
%% \bibliographystyle{JHEP}
%% \bibliography{biblio.bib}

%% or
%% [B] Manual formatting (see below)
%% (i) We suggest to always provide author, title and journal data or doi:
%% in short all the informations that clearly identify a document.
%% (ii) please avoid comments such as "For a review'', "For some examples",
%% "and references therein" or move them in the text. In general, please leave only references in the bibliography and move all
%% accessory text in footnotes.
%% (iii) Also, please have only one work for each \bibitem.
\bibliography{biblio}
\bibliographystyle{JHEP.bst}

\end{document}